\documentclass[useAMS,usenatbib]{mn2e}
\usepackage{graphicx}
\usepackage{longtable}
\title[A quasar showing remarkably variable BALs]{SDSS J1138+3517: A quasar showing remarkably variable broad absorption lines}
\author[C. Wildy, M. R. Goad and J. T. Allen]{C. Wildy$^{1}$\thanks{E-mail:
    cw268@le.ac.uk}, M. R. Goad$^{1}$ and
  J. T. Allen$^{2}$\\ $^{1}$University of Leicester, Department of
  Physics and Astronomy, University Road, Leicester, UK\\ $^{2}$Sydney
  Institute for Astronomy, School of Physics, A28, The University of
  Sydney, NSW 2006, Australia}

  \renewenvironment{thebibliography}[1]{%
    \begin{oldthebibliography}{#1}%
      \setlength{\parskip}{0ex}%
      \setlength{\itemsep}{0ex}%
  }%
  {%
    \end{oldthebibliography}%
  }

\begin{document}

\date{Accepted}

\pagerange{\pageref{firstpage}--\pageref{lastpage}} \pubyear{0000}

\maketitle

\label{firstpage}

\begin{abstract}

\noindent We report on the highly variable Si~{\sc iv} and C~{\sc iv} broad absorption lines in SDSS J113831.4+351725.2 across four observational epochs. Using the Si {\sc iv} doublet components, we find that the blue component is usually saturated and non-black, with the ratio of optical depths between the two components rarely being 2:1. This indicates that these absorbers do not fully cover the line-of-sight and thus a simple apparent optical depth model is insufficient when measuring the true opacity of the absorbers. Tests with inhomogeneous (power-law) and pure-partial coverage (step-function) models of the absorbing Si {\sc iv} optical depth predict the most un-blended doublet's component profiles equally well. However, when testing with Gaussian-fitted doublet components to all Si~{\sc iv} absorbers and averaging the total absorption predicted in each doublet, the upper limit of the power law index is mostly unconstrained. This leads us to favour pure partial coverage as a more accurate measure of the true optical depth than the inhomogeneous power law model. 

The pure-partial coverage model indicates no significant change in covering fraction across the epochs, with changes in the incident ionizing flux on the absorbing gas instead being favoured as the variability mechanism. This is supported by (a) the coordinated behaviour of the absorption troughs, (b) the behaviour of the continuum at the blue end of the spectrum and (c) the consistency of photoionization simulations of ionic column density dependencies on ionization parameter with the observed variations. Evidence from the simulations together with the C~{\sc iv} absorption profile indicates that the absorber lies outside the broad line region, though the precise distance and kinetic luminosity are not well constrained.

\end{abstract}


\section{Introduction}

Blue-shifted quasar absorption lines, which indicate the presence of large amounts of outflowing material from the central engine, have become an increasingly important topic over the past two decades. This followed the realisation of the potential influence these flows can exert on the host galaxy by restricting black hole growth \citep{springel05,king10} and contributing to the evolution of galactic bulges \citep{magorrian98,silk98}. This feedback process is also supported by evidence that outflows could extend hundreds of parsecs into the surrounding interstellar medium \citep{barlow94}, depositing significant amounts of kinetic energy in the process \citep{arav13}.

Absorption features observed in the ultraviolet (UV) portion of the quasar's rest-frame spectrum can be blue-shifted by as much as 0.1c, where $c$ is the speed of light in a vacuum, from the corresponding emission line centre. These absorption lines are divided into categories according to their velocity width, the widest being broad absorption lines (BALs) extending over at least 2000 km s$^{-1}$ where flux is less than 90 per cent of the continuum \citep{weymann91}. Quasars hosting at least one BAL are known as broad absorption line quasars (BALQSOs). Absorption can also be seen as narrow absorption lines (NALs), which extend over a few hundred km s$^{-1}$ and mini-BALs, which are of intermediate width between NALs and BALs \citep{hamann04}.

The population of BALQSOs, of which SDSS J113831.4+351725.2 (hereafter SDSS J1138+3517) is a member, account for approximately 10 per cent to 20 per cent of the total quasar population \citep{reichard03a,knigge08,scaringi09}. However SDSS target selection effects may make this value as high as 41 per cent \citep{allen11}. If BAL observability is governed by line of sight orientation, then large-scale outflows may occur in the vast majority of quasars, even where no BALs are visible \citep{schmidt99}. Common BAL subtypes include High-Ionization BALQSOs (HiBALs) ($\sim$85 per cent) whose spectra only show broad absorption due to high ionization transitions such as C~{\sc iv} $\lambda$$\lambda$1548,1550 Si~{\sc iv} $\lambda$$\lambda$1394,1403 and N~{\sc v} $\lambda$$\lambda$1239,1243 \citep{sprayberry92,reichard03b}. Low Ionization BALQSOs (LoBALs), in addition to high-ionization BALs, also exhibit broad absorption resulting from low ionization species such as Mg~{\sc ii}, Al~{\sc iii} and on rare occasions Fe~{\sc ii} (known as FeLoBALs).

Ionizing continuum changes are known to drive broad emission line (BEL) variability \citep{peterson98,vandenberk04,wilhite06}, however there is as yet no consensus behind the dominant mechanism involved in absorption line variablity. Several large statistical studies of BAL variability have been published in recent years \citep{lundgren07,gibson08,gibson10,capellupo11,capellupo12,capellupo13,filizak13,wildy14}. These studies refer to coordinated variability across widely separated absorption troughs as evidence favouring changes in input ionizing flux as the main mechanism behind variability \citep{filizak13}, while variability over narrow velocity ranges within troughs is attributed to absorbing gas moving across the line of sight \citep{gibson08,capellupo12}. 

As in \citet{wildy14}, we use a novel method to reconstruct the un-absorbed spectral profile \citep{allen11} against which the absorption can be measured.  Absorption is identified based on this emission profile as it contains the total input spectral flux (continuum+emission line) entering the absorbing gas at each wavelength bin. We use this method to estimate the column density at each epoch for the Si~{\sc iv} and C~{\sc iv} ions. This has advantages over purely continuum-based absorber strength measurements as it allows absorption depth to be accurately identified in spectral regions where the absorption is overlapping with an emission line profile, allowing absorption features to be studied to arbitrarily low (less negative) velocities.

\subsection{The Unusual BAL behaviour of SDSS J1138+3517}

The quasar examined in this paper, SDSS J1138+3517, has a redshift of z=2.122 \citep{hewett10} and was investigated as part of the multi-BALQSO variability study undertaken in the rest-frame UV by \citet{wildy14}. Using a sample of 50 quasars, Si~{\sc iv} $\lambda$1400 and C~{\sc iv} $\lambda$1549 BALs were identified and statistics gathered on their equivalent width (EW) changes over 2 epochs separated by quasar rest-frame timescales ranging from approximately 1 to 4 years. Of the 50 BALQSOs, SDSS J1138+3517 showed the largest absolute change in EW in both Si~{\sc iv} and C~{\sc iv} absorption as well as a large fractional EW change ($|$$\Delta$$EW$$/$$\langle$$EW$$\rangle$$|$), occurring over a rest-frame time interval of $\sim$1 year. A third epoch indicated another dramatic variation in the BALs, while a fourth epoch showed relatively little change from the third except for significant variability in one C~{\sc iv} absorber, with these subsequent observations also being made in successive 1-year rest-frame time intervals.  

In \citet{filizak13}, the variability of 428 C~{\sc iv} BALs in 291 quasars was found to produce a $\Delta$$EW$ dependence on rest-frame time interval which is well predicted by a random-walk model (see Fig.~29 of that paper). Using the same BAL identification scheme as that study, the variation across the two observational epochs examined in \citet{wildy14} of the most variable C~{\sc iv} BAL region in SDSS J1138+3517 (spanning $-$23\,400 km s$^{-1}$ to $-$7500 km s$^{-1}$) is $\Delta$$EW$=$-$13.8$\pm$2.9 \AA{}. This value lies well outside the random-walk model prediction at the epoch separation time-interval of 360 days in the quasar rest-frame, suggesting that this object may exhibit a variability mechanism which is rare in the BALQSO population.

In \citet{filizak14}, which uses the same BAL identification scheme as above, the variability of C~{\sc iv} BALs was investigated over timescales ranging from $\sim$1 to a few years, including those in quasars additionally containing Si~{\sc iv} BALs of overlapping velocity. Again examining the most variable C~{\sc iv} BAL in SDSS J1138+3517 identified using their definition, the value of $\Delta$$EW$, along with a value of $\Delta$$EW$$/$$\langle$$EW$$\rangle$=$-$1.3$\pm$0.3 would place it among the top few most variable C~{\sc iv} BALs in their sample of 454 (includes C~{\sc iv} BALs both unaccompanied by Si~{\sc iv} absorption and overlapping in velocity with a Si~{\sc iv} BAL), highlighting the extreme variability of this source. In addition, an investigation by \citet{gibson10} included observations of two quasars over a similar rest-frame time interval of between 0.5 and 2 years, neither of which showed variability as dramatic as in SDSS J1138+3517.

Answering the question of what drives the strong BAL variability observed in this quasar is the principal objective of this investigation. Observations of the source are detailed in Section 2. In Section 3, we quantify Si~{\sc iv} and C~{\sc iv} ionic column densities using different assumptions regarding the line-of-sight geometry, including models where there is incomplete coverage of the emission region by a homogeneous absorber. This is aided by the use of Gaussian profiles to fit the Si~{\sc iv} absorption components, which allows calculations to be performed based on the changes in these components even where they are not fully un-blended. In Section 4 we compute a grid of photoionization models, allowing examination of the theoretical variation of ionic column density over a range of ionization parameter at given hydrogen number densities and column densities. These grid models are compared to the best results for the ionic column densities from Section 3, allowing us to test the hypothesis of changes in ionizing flux being the driver of BAL variability. In Section 5 we discuss the results from Sections 3 and 4 and their implications for the variability mechanism. We also attempt to constrain the mass outflow rate and hence kinetic luminosity. 

This paper assumes a flat $\Lambda$CDM cosmology with H$_{0}$=67 km s$^{-1}$ Mpc$^{-1}$, $\Omega_{M}$=0.315 and $\Omega_{\Lambda}$=0.685, as reported from the latest results of the Planck satellite \citep{planckxvi14}.

\section{Observations}

	  \subsection{The Sloan Digital Sky Survey}

The epoch 1 observation was obtained from DR6 of the Sloan Digital Sky Survey (SDSS). Beginning in 2000, the SDSS had imaged approximately 10\,000 deg$\sp{2}$ of the sky as of June 2010 \citep{schneider10},
using the 2.5m Telescope at the Apache Point Observatory, New Mexico, USA \citep{sdss}. Imaging is carried out using a CCD camera that operates in drift-scanning mode \citep{gunn98}. Five broad band
filters, u,g,r,i and z, normalised to the AB system, are used, covering a wavelength range of 3900 to 9100 \AA{} with an approximate spectral resolution of $\lambda$/$\Delta$$\lambda$=2000 at 5000~\AA{}
\citep{stoughton02}. Further information on the SDSS data reduction process is given in \citet{lupton01} and \citet{stoughton02}.

	  \subsection{The William Herschel Telescope}

The William Herschel Telescope (WHT) ISIS (Intermediate dispersion Spectrograph and Imaging System) double-armed (red and blue) spectrograph was used for longslit spectroscopy of the target during semester A 2008, 2011 and 2014 corresponding to epochs 2,3 and 4 respectively, with epochs 3 and 4 being observed as part of the ING service programme and epoch 2 taken from the sample used by \citet{cottis10}. The observations provide spectral coverage from 3000 to 10\,000 \AA{}. For epoch 2 the 5700 dichroic was used, providing spectra from the two arms with overlap in the range 5400 to 5700 \AA{}. The R158B and R158R gratings, giving a nominal spectral resolution of 1.6\AA{} per pixel in the blue and 1.8\AA{} per pixel in the red, were also utilised for epoch 2. Epoch 3 and 4 observations employed the 5300 dichroic, giving an overlap range of 5200-5500 \AA{}, while the R300B and R316R gratings were used for the blue and red observations, providing spectral resolutions of 0.86 \AA{} per pixel and 0.93 \AA{} per pixel respectively. In the dichroic overlap regions, an error weighted mean was used to combine the spectra from the two arms for each observation. For all WHT observations, a slit width of 1.5" was chosen to match typical seeing conditions and give reasonable throughput without compromising the spectral resolution. Multiple exposures were taken (in order to remove cosmic ray hits) bracketed by arc-lamp exposures for wavelength calibration. Standard star spectra were taken on the observing nights to allow flux calibration of the targets.  Correction of CCD images using bias and flat field exposures was performed within the {\sc iraf} \emph{ccdproc} routine. Spectra were extracted, along with sky background removal, using the \emph{apall} task within the {\sc iraf} \emph{longslit} package. This used an extraction slit width of 10 pixels and optimal weighting, with 2 pixels per wavelength bin. Wavelength and flux calibration were applied to the extracted spectra within this same package. The spectra were placed on an absolute flux scale using spectra of a photometric standard observed at similar airmass. No attempt was made, such as through the use of grey shifts, to adjust the flux scale on the calibrated spectra towards levels which more closely matched across the epochs. Table~\ref{tab:obs} summarises the SDSS J1138+3517 observations.

\begin{table*}
\begin{center}
\caption{Details of observations obtained of SDSS J1138+3517}
\begin{tabular}{l l l l l l l l}
\hline Epoch&Observation Date&$\Delta$$t$$_{qrest}$&Type&Exp. time&Grating&Pixel Resolution (5000 \AA{})&Mean Airmass\\
 & &(days)& &(s)& &(km s$^{-1}$)& \\
\hline
1&03 May 2005&0&SDSS&6$\times$2400&Red \&{} Blue&150&1.01\\
2&27 May 2008&359&WHT&3$\times$1200&R158B \&{} R158R&108&1.14\\
3&29 May 2011&710&WHT&2$\times$1800&R300B \&{} R316R&56&1.80\\
4&09 May 2014&1053&WHT&2$\times$1800&R300B \&{} R316R&56&1.06\\
\hline
\end{tabular}
\label{tab:obs}
\end{center}
$\Delta$$t$$_{qrest}$ is quasar rest-frame time interval since Epoch 1\\
\end{table*}

\section{Analysis}

\subsection{Spectral Properties}

The observed spectra at each epoch correspond to rest-frame UV wavelengths which include the region containing Si~{\sc iv} and C~{\sc iv} emission and the corresponding blueshifted absorption lines resulting from those same atomic transitions in outflowing gas. Since the spectral resolution differs between spectra, in order to accurately identify changes in absorption between each epoch the SDSS and WHT epoch 3 and epoch 4 spectra were convolved with Gaussians of appropriate full-width half maxima to approximate the resolution of the WHT epoch 2 observation (using skyline widths as a guideline), followed by resampling of all 4 spectra onto a wavelength grid of 3.7 \AA{}. This closely matches the bin width of the epoch 2 observation. Small differences in wavelength calibration were removed by aligning narrow emission and absorption features in the WHT spectra with the same lines in the SDSS spectrum. The SDSS and both WHT spectra were corrected for the effects of Milky Way Galactic Extinction using the method of \citet{cardelli89} with A$_{V}$ values taken from \citet{schlegel98}.

Using the reconstruction of the unabsorbed SDSS spectrum from \citet{wildy14} originally developed in \citet{allen11}, appropriate reconstructions were also generated for the three WHT observations. This was achieved as follows: (i) A power-law continuum was fitted to each spectrum using line-free continuum bands as outlined in \citet{wildy14}, (ii) The SDSS continuum was subtracted from the SDSS reconstruction, (iii) The WHT spectra had their respective continua subtracted, (iv) In the continuum subtracted SDSS reconstruction, the Si~{\sc iv} $\lambda$1400, C~{\sc iv} $\lambda$1549, O~{\sc i} $\lambda$1304 and C~{\sc ii} $\lambda$1334 emission line boundaries were identified, (v) The emission lines from part (iv) were scaled to match the red side (unabsorbed) of the emission lines in each of the WHT observations, and (vi) The continuum was added to the scaled emission lines for each of the WHT observations, creating appropriate reconstructions for epochs 2,3 and 4. 

\begin{figure}
\centering
\resizebox{\hsize}{!}{\includegraphics[angle=0,width=8cm]{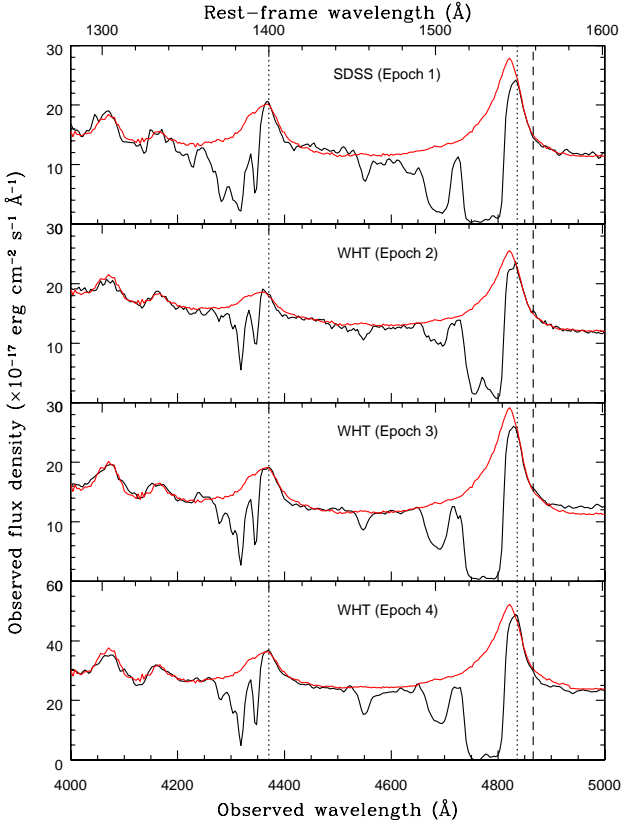}}
\caption{Spectra for epochs 1 to 4 of SDSS J1138+3517, with top panel to bottom panel in order of observation date starting with the earliest. The observed spectrum is in black while the reconstruction of the unabsorbed spectrum is in red. Vertical dotted lines indicate the laboratory wavelengths of the Si~{\sc iv} and C~{\sc iv} emission lines. The vertical dashed line indicates the maximum wavelength used in calculating the C~{\sc iv} emission scaling, emission longward of this point is not accurately reconstructed.}
\label{fig:recs}
\end{figure}

For the SDSS reconstruction used in the subsequent analysis, the procedure was carried out to scale the Si~{\sc iv} emission line, as the original reconstruction underestimates the Si {\sc iv} emission (the reconstructions were optimised for C~{\sc iv} as described in \citet{allen11}). The region used for calculating the appropriate C~{\sc iv} line scaling is restricted to the interval between line centre and an observed wavelength of 4866 \AA{}. This is to avoid the He~{\sc ii}+O{\sc iii} absorption and emission complex immediately longward of C~{\sc iv}. The Si~{\sc iv} emission line was shifted by two wavelength bins blueward in the epoch 2 reconstruction relative to the SDSS reconstruction to match the shifted peak of the emission line in the epoch 2 spectrum. Based on the synthetic BAL based error estimation method described in Section 3 of \citet{wildy14}, the fractional error on the reconstruction for a quasar of this r-band S/N and redshift is estimated to be 5 per cent of the flux value at each resolution element for all calculations described in this paper. The observed spectra for all 4 epochs (solid black lines), as well as their final reconstructions (solid red lines), are illustrated in Fig.~\ref{fig:recs}.

\begin{figure}
\centering
\resizebox{\hsize}{!}{\includegraphics[angle=0,width=8cm]{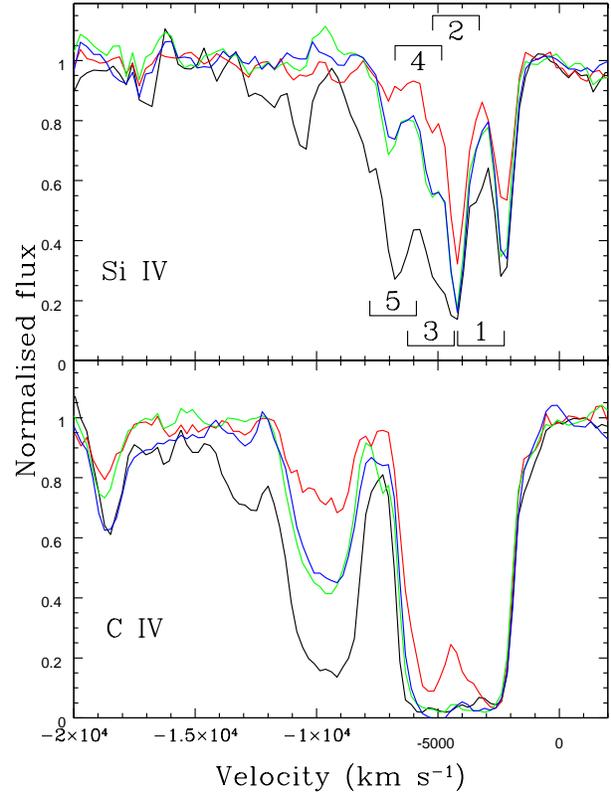}}
\caption{Si~{\sc iv} (upper panel) and C~{\sc iv} (lower panel) absorption regions in red component velocity space. Epoch 1 (SDSS) spectra are in black, epoch 2 (1st WHT) spectra are in red, epoch 3 (2nd WHT) spectra are in green and epoch 4 (3rd WHT) spectra are in blue. Spectra are normalised to the un-absorbed reconstruction. The Si~{\sc iv} Gaussian components (described in Section 3.3) are numerically labelled in order of increasing outflow velocity}
\label{fig:ionspec}
\end{figure}

The dramatic variability in both ions' absorbers across the first three epochs is clear when the spectra are normalised to the reconstructions. The transitions of interest are both doublets, with rest-frame laboratory wavelengths of 1393.76 and 1402.77 \AA{} contributing to Si~{\sc iv} $\lambda$1400 and correspondingly 1548.20 and 1550.77 \AA{} for C~{\sc iv} $\lambda$1549. Relative to the laboratory-frame rest-wavelength of the red component of each doublet, absorption is significant between 0 and $\sim$$-$13\,000 km s$^{-1}$ for Si~{\sc iv} and between 0 and $\sim$$-$20\,000 km s$^{-1}$ for C~{\sc iv}, where a negative value indicates outflowing material (blueshifted). As can be seen in Fig.~\ref{fig:ionspec}, the C~{\sc iv} troughs are deeper than the Si {\sc iv} troughs at similar velocities. The deepest troughs show the least variability and are located at the lowest (least negative) velocities, as has been reported in previous BAL studies \citep{lundgren07,capellupo11}.

Due to its relative lack of variation from the previous observation, especially in the Si~{\sc iv} absorption lines where identifiable doublets are present, epoch 4 is left out of the analysis in subsequent parts of Section 3 and Section 4. Epoch 4 is instead discussed further in Section 5.3.

\subsection{Lower Limits for Outflow Column Densities using Direct Integration}

Minimum values for both the Si~{\sc iv} and C~{\sc iv} column densities can be estimated by assuming unsaturated absorbers completely cover the emitting line+continuum region along our line of sight with a constant optical depth across the tangential plane \citep{savage91}. The C~{\sc iv} and Si~{\sc iv} transitions have a doublet structure with known oscillator strengths. The total column density of the ion contributing to a given doublet component may be calculated according to

\begin{equation}
\label{eqn:dicalc}
N_{ion}=\frac{m_{e}c}{\pi{}e^{2}f\lambda{}}\int \tau{}\left(v\right)dv
\end{equation}

\noindent where $N_{ion}$ is the ionic column density, $v$ is the velocity relative to the laboratory rest-frame wavelength, $m_{e}$ is the electron mass, $c$ is the speed of light, $e$ is the elementary charge, $f$ is the oscillator strength, $\lambda$ is the laboratory rest-frame wavelength and $\tau(v)$ is the optical depth at a given velocity. At a redshift of $z=2.122$, the doublet separation at zero velocity is approximately 28.1 \AA{} for Si~{\sc iv} and 8.0 \AA{} for C~{\sc iv}. As the C~{\sc iv} separation is only slightly greater than two wavelength bins, it is practically impossible, given their large intrinsic widths, to identify individual C~{\sc iv} components. Though some Si~{\sc iv} components are well-separated, many of the doublets are blended to various extents with neighbouring absorption features. However, the optical depth of overlapping components will be the sum of the individual components and the optical depth of the red component is known to be half that of the blue component for both ions, given the ratio of their oscillator strengths. Therefore, by integrating over all of the absorption for a given transition using the value of $f$ corresponding to the red component, a minimum value of $N_{ion}$ can be found by taking 1/3 of the calculated value. The velocity ranges integrated over and resulting estimated ionic column densities are shown in Table~\ref{tab:ditab}.

\begin{table*}
\begin{center}
\caption{C~{\sc iv} and Si~{\sc iv} velocity limits and ionic column densities for each epoch using direct integration of the absorption profile.}
\begin{tabular}{l l l l l}
\hline Transition&Velocity Limit$^{\dagger}$&Epoch 1 Ionic Column Density&Epoch 2 Ionic Column Density&Epoch 3 Ionic Column Density\\
 &(km s$^{-1}$)&($\times$10$^{14}$cm$^{-2}$)&($\times$10$^{14}$cm$^{-2}$)&($\times$10$^{14}$cm$^{-2}$)\\
\hline
Si~{\sc iv} $\lambda$1400&-13\,100&24.7$\pm$2.66&7.21$\pm$2.37&11.7$\pm$2.41\\
C~{\sc iv} $\lambda$1549&-19\,700&190$\pm$15.2&95.8$\pm$8.64&151$\pm$11.3\\
\hline
\end{tabular}
\label{tab:ditab}
\end{center}
$^{\dagger}$Integration spans zero velocity to velocity limit\\
\end{table*}

Since the C~{\sc iv} doublet is unresolved, this method is the only means of estimating a lower limit for the column density of this ion. For Si~{\sc iv}, methods involving modelling the components need to be checked for consistency with the lower limit derived here.

\subsection{Gaussian Components of the Si IV Outflow}

\begin{figure}
\centering
\resizebox{\hsize}{!}{\includegraphics[angle=0,width=8cm]{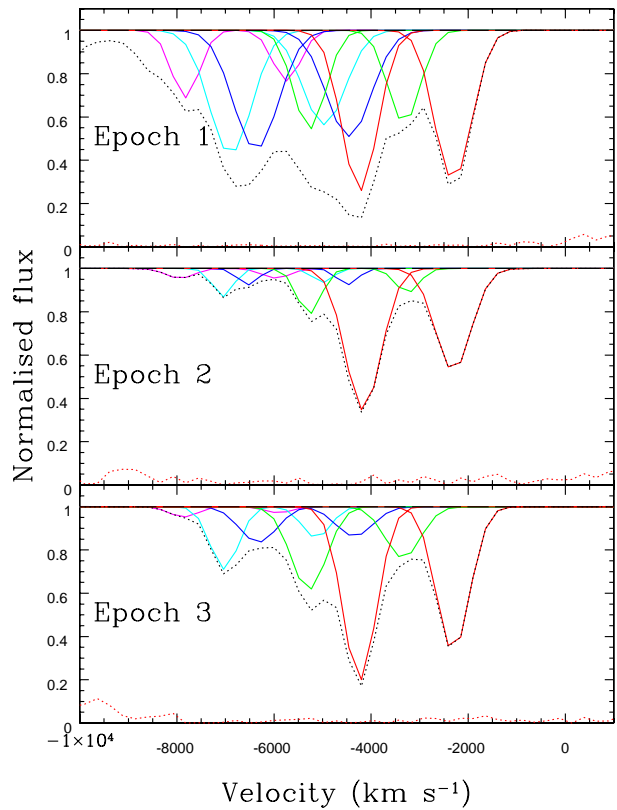}}
\caption{Epoch 1 (top panel), epoch 2 (middle panel) and epoch 3 (lower panel) models of Si~{\sc iv} doublet absorption lines in red component velocity space (related doublet components are shown in the same colour). The black dotted line describes the total model absorption profile, while the red dotted line near the bottom of each panel indicates the difference between the model profile and the observed profile.}
\label{fig:si4comps}
\end{figure}

The Si~{\sc iv} doublet velocity separation is $\sim$1920 km s$^{-1}$, allowing some individual components to be identified. Assuming narrow components of quasar absorption lines follow an approximately Gaussian profile, e.g. \citet{hamann11}, model profiles can be constructed for the doublet lines. Spectral fitting is performed using the \emph{specfit} package within {\sc iraf}, requiring three free parameters for each Gaussian component, namely the wavelength at line centre, line full width half maximum (FWHM) and line EW. Appropriate restrictions are applied to these parameters, i.e. the EW ratio between the blue and red components of the doublet must be between 2:1 and 1:1, the difference in velocities at line centre must not differ from the expected doublet separation by more than one velocity bin, and the widths must be the same. By using an initial 'guess' for these parameters' values, spectral fitting is performed using chi-square minimisation. In total, five Si~{\sc iv} doublets are identified across the 3 epochs, spanning the range $-$10\,000$\leq$$v$$\leq$0 km s$^{-1}$. These are shown in Fig.~\ref{fig:si4comps}.

Comparing the Gaussian fit to the almost un-blended red component of the lowest velocity doublet in epoch 2 (shown in red in Fig.~\ref{fig:si4comps}) to the data points at these velocities gives a fractional error in the fit of 0.033. To be conservative a larger error of 5 per cent on each point of all Gaussian profiles is assumed for calculations involving these model profiles. While the EW and occasionally the FWHM change from one epoch to the next, there is no significant change in the line centre velocity and hence no evidence for acceleration of the outflowing material. This lack of acceleration has been noted in previous studies \citep{hamann08,hidalgo11}. The best-fit parameters obtained by \emph{specfit} for each doublet component are provided in Table~\ref{tab:si4tab}.

\begin{table*}
\begin{center}
\caption{List of components with parameters for Gaussian model profiles. Doublets are listed in order of increasingly negative velocity (1 is lowest, 5 is highest). Doublet velocity is measured from the red component rest wavelength.}
\begin{tabular}{l l ll ll ll}
\hline Absorber&Velocity$^{\ddagger{}}$&\multicolumn{2}{c}{Epoch 1}&\multicolumn{2}{c}{Epoch 2}&\multicolumn{2}{c}{Epoch 3}\\
 &(km s$^{-1}$)& & & & & & \\
\hline
 & &FWHM$^{\ddagger{}}$&EW$^{\dagger{}}$&FWHM$^{\ddagger{}}$&EW$^{\dagger{}}$&FWHM$^{\ddagger{}}$&EW$^{\dagger{}}$\\
 & &(km s$^{-1}$)&(\AA{})&(km s$^{-1}$)&(\AA{})&(km s$^{-1}$)&(\AA{})\\
\hline\hline
1 (Red)&$-$2300&900&3.18$\pm$0.09&900&2.03$\pm$0.07&800&2.69$\pm$0.04\\
1 (Blue)& & &3.18$\pm$0.16& &2.80$\pm$0.07& &3.21$\pm$0.05\\
\hline
2 (Red)&$-$3300&800&1.65$\pm$0.12&600&0.32$\pm$0.07&900&1.01$\pm$0.06\\
2 (Blue)& & &1.76$\pm$0.08& &0.60$\pm$0.08& &1.62$\pm$0.07\\
\hline
3 (Red)&$-$4300&1100&2.62$\pm$0.24&500&0.19$\pm$0.08&900&0.60$\pm$0.11\\
3 (Blue)& & &2.92$\pm$0.18& &0.19$\pm$0.07& &0.72$\pm$0.07\\
\hline
4 (Red)&$-$4800&1000&2.21$\pm$0.22&500&0.16$\pm$0.08&700&0.48$\pm$0.07\\
4 (Blue)& & &2.88$\pm$0.18& &0.21$\pm$0.08& &0.97$\pm$0.06\\
\hline
5 (Red)&$-$5900&700&0.77$\pm$0.15&700&0.15$\pm$0.08&700&0.09$\pm$0.06\\
5 (Blue)& & &1.02$\pm$0.13& &0.15$\pm$0.11& &0.16$\pm$0.06\\
\hline\hline
\end{tabular}
\label{tab:si4tab}
\end{center}
$^{\dagger{}}$EW in quasar rest-frame\\
$^{\ddagger{}}$Error $\sim$200 km s$^{-1}$\\
\end{table*}

Upon examination of an individual component, the following equation can be used to estimate the column density giving rise to the doublet to which it belongs: 

\begin{equation}
\label{eqn:pod}
N=\frac{m_{e}cb}{\sqrt{\pi{}}e^{2}f\lambda{}}\tau_{0}
\end{equation}

\noindent where $\tau$$_{0}$ is the peak optical depth and $b$ is the Doppler parameter (related to line width by $b=\sqrt{2}\times{}\rm{FWHM}/2.355$). This assumes a Voigt profile for the absorbers, which tends towards the Gaussian case in the limit of low optical depth. Red components are less likely to be affected by saturation since they are weaker than their corresponding blue components. As a result, column density measurements are taken from red doublet lines. As in Section 3.2, estimates of the absorbing column density are lower limits as the absorber may not cover 100 per cent of the emission region. Summing the values obtained for the five doublets listed in Table~\ref{tab:si4tab} gives estimates of the total column densities as $N_{\rm Si~{IV},1}$=33.5$\pm$1.49, N$_{\rm Si~{IV},2}$=7.79$\pm$0.50 and N$_{\rm Si~{IV},3}$=14.7$\pm$0.47 in units of 10$^{14}$ cm$^{-2}$ for the first, second and third epochs respectively. As expected, these are larger than the limits in Table~\ref{tab:ditab} since blue:red doublet component ratios are less than 2:1.

\subsection{A Pure Partial Coverage Model}

Many absorption line systems have been found to be best described by a situation in which material of a certain optical depth covers a fraction of the emission source, leaving the rest of the source uncovered \citep{arav99,dekool01} otherwise known as a pure partial coverage (PPC) model. Under these conditions the apparent optical depth $\tau=ln(1/I_{res})$, where $I_{res}$ is the residual intensity (normalised to the unabsorbed continuum+line emission flux), will underestimate the true optical depth of the absorbing gas and consequently will underestimate the column density. The observed flux $I_{app}$ is related to the covering fraction $C$ by $I_{app}=CI_{out}+(1-C)I_{0}$, where $I_{0}$ is the unabsorbed emission flux and $I_{out}$ is the true output flux from the absorber. Thus the covering fraction $C$ as a function of velocity is given by

\begin{equation}
\label{eqn:cfrac}
C\left(v\right)=\frac{I^{2}_{r}\left(v\right)-2I_{r}\left(v\right)+1}{I_{b}\left(v\right)-2I_{r}\left(v\right)+1}
\end{equation}

\noindent where I$_{r}\left(v\right)$ and I$_{b}\left(v\right)$ are the apparent residual fluxes of the red and blue components respectively at a given velocity $v$. Since the true optical depth $\tau=ln(I_{0}/I_{out})$, the following is also true

\begin{equation}
\label{eqn:pcod}
\tau{}=-ln\left(\frac{I_{r}-I_{b}}{1-I_{r}}\right)
\end{equation}

\noindent where $\tau$ is the optical depth of the red component at a specific velocity. Since $C$$\leq$1, the values of I$_{r}$ and I$_{b}$ can be shown to be constrained to the range $I_{r}\geq{}I_{b}\geq{}I_{r}^{2}$. Assuming this condition is met, it is then straightforward to calculate the column density using Equation~\ref{eqn:dicalc}. 

\begin{figure}
\centering
\resizebox{\hsize}{!}{\includegraphics[angle=0,width=8cm]{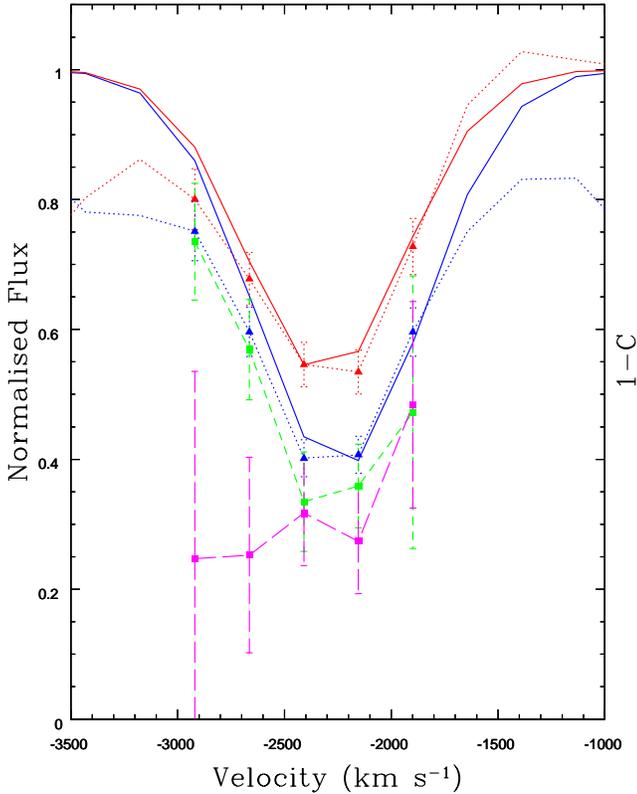}}
\caption{The absorber 1 profiles of the red and blue components interpolated onto the red component's velocity grid. The solid red and blue lines represent the Gaussian models, the dotted lines represent the observed fluxes, the green short-dashed line indicates the $1-C$ values at these points and the magenta long-dashed line represents the e$^{-\tau}$ values of the red component. Appropriately coloured triangles and squares represent the observed points in velocity space. While the $1-C$ profile is a good match for the blue absorber, the e$^{-\tau}$ profile does not match the red profile.}
\label{fig:intp}
\end{figure}

Evidence for pure partial covering can be obtained by comparing the residual flux velocity dependence of a (almost) saturated absorption line to the profile of $1-C$. Similar profiles would suggest the shape of the absorption profile is determined by the velocity-dependent covering fraction of an opaque absorber rather than intrinsic differences in optical depth. An additional test is to compare the profile of e$^{-\tau}$ to the red component of an absorption doublet. If the profiles do not match, the absorption line profile cannot be due to an absorber which is homogeneous and completely covers the emission region. In order to carry out these tests, a doublet with unblended components is desirable. The closest example available in the dataset is absorber 1 in the epoch 2 spectrum. From the values of the EWs of the red and blue components (see Table~\ref{tab:si4tab}), the blue:red ratio is less than 2:1, indicating that the blue component is probably saturated. After placing the blue component onto the red component velocity grid by interpolation, six velocity bins are found which follow the general shape of the corresponding Gaussian models for both the red and blue components. Five of these points satisfy I$_{r}$$\geq$I$_{b}$$\geq$I$_{r}^{2}$ and are illustrated in Fig.~\ref{fig:intp}.

From the profile traced by $1-C$ (Fig.~\ref{fig:intp}, dashed green line), it is clear that the saturated blue component could have its profile determined by the uncovered fraction. It is also obvious that the residual intensity of the red component is a poor match to $e^{-\tau}$. Together these provide \emph{strong evidence that partial coverage is a significant effect in this absorber} and possibly others in this quasar. 

Given that this is the only doublet where meaningful results can be obtained on an individual velocity bin basis, and even in this case it can only be done close to the peak of the absorption components, it is necessary to use the Gaussian models to achieve estimates of column density across a greater proportion of the Si~{\sc iv} absorption for the 3 epochs. Individual velocity bins in the Gaussian models have no physical realisation, so unlike for the previous example it is impossible to construct $1-C$ profiles using Equation~\ref{eqn:cfrac}. However if for each doublet component, model fluxes are averaged over all velocities spanned by the absorption, individual components can be treated as individual velocity bins. This method can subsequently be used to estimate average values of covering fraction and true optical depth for each doublet. Using this true optical depth in Equation~\ref{eqn:dicalc} gives the doublet's column density along the line of sight when $N_{ion}$ is multiplied by the covering fraction. These values are listed in Table~\ref{tab:ppcabs}. Due to the weakness of absorbers 3 to 5 in epochs 2 and 3 this can only be performed across all absorbers in epoch 1, since at small absorber depths optical depth and covering fraction measurements are unreliable due to their sensitive dependence on doublet component ratio.

\begin{table*}
\begin{center}
\caption{Table of covering fractions and column densities calculated for absorbers 1 to 5. Absorbers 3 to 5 are not listed for epochs 2 and 3 due to their low strength.}
\begin{tabular}{l ll ll ll}
\hline Absorber&\multicolumn{2}{c}{Epoch 1}&\multicolumn{2}{c}{Epoch 2}&\multicolumn{2}{c}{Epoch 3}\\
\hline
 &$C$&$N_{ion}$&$C$&$N_{ion}$&$C$&$N_{ion}$\\
 & &$\times$10$^{14}$cm$^{-2}$& &$\times$10$^{14}$cm$^{-2}$& &$\times$10$^{14}$cm$^{-2}$\\
\hline\hline
1&0.225$\pm$0.013&23.3$^{+15.6}_{-14.0}$&0.234$\pm$0.042&7.17$^{+3.56}_{-2.87}$&0.236$\pm$0.020&12.3$^{+4.40}_{-3.881}$\\
2&0.187$\pm$0.017&11.0$^{+8.68}_{-7.40}$&0.318$^{+0.682}_{-0.318}$&0.773$^{+13.7}_{-0.773}$&0.275$\pm$0.165&2.898$^{+4.010}_{-0.165}$\\
3&0.339$\pm$0.017&14.1$^{+4.20}_{-3.88}$& & & & \\
4&0.366$\pm$0.032&8.45$^{+2.34}_{-2.09}$& & & & \\
5&0.122$\pm$0.043&2.94$^{+3.61}_{-2.27}$& & & & \\
\hline\hline
\end{tabular}
\label{tab:ppcabs}
\end{center}
\end{table*}

\subsection{An Inhomogeneous Absorber Model}

An alternative to the PPC model is a situation whereby the optical depth varies with spatial location across the line of sight. This was originally developed using a power-law dependent optical depth model in \citet{dekool02} and was subsequently investigated by \citet{arav05}, who found power-law and Gaussian shaped optical depth dependencies on a 1-dimensional spatial location to be inadequate in describing O~{\sc vi} absorbers in Mrk 279. Here, we attempt a similar investigation using a power-law model of the form $\tau$(x,$\lambda$)=$\tau$$_{max}$($\lambda$)x$^{a}$, where $\tau$(x,$\lambda$) is the optical depth at position $x$ on the surface of the absorber and at wavelength $\lambda$, $\tau$$_{max}$($\lambda$) is the maximum optical depth of the absorber and $a$ is the power-law index. 

Following \citet{arav05}, we make two simplifying assumptions to aid calculations without any loss of generality. First, we assume that the surface brightness is uniform across the face of the emission source such that S(x,y,$\lambda$)=1, where x and y are spatial positions in the plane of the line of sight and S is arbitrarily normalised. We also simplify the optical depth spatial dependence by using a 1-dimensional model, so that $\tau$(x,y,$\lambda$)=$\tau$(x,$\lambda$). The $x$ values are fixed to span an interval 0$\leq$x$\leq$1 for simplicity. The observed residual flux at a given velocity bin will then be found using $I_{res}$=$\int_{0}^{1}$$e^{-\tau(x)}$$dx$. To assess a broad range of power-law indices, values from 0 to 10 with step size of 0.1 are tested. In order to integrate over the $x$ range, it is effectively divided into 1000 bins, giving $dx$=0.001.

\begin{figure}
\centering
\resizebox{\hsize}{!}{\includegraphics[angle=0,width=8cm]{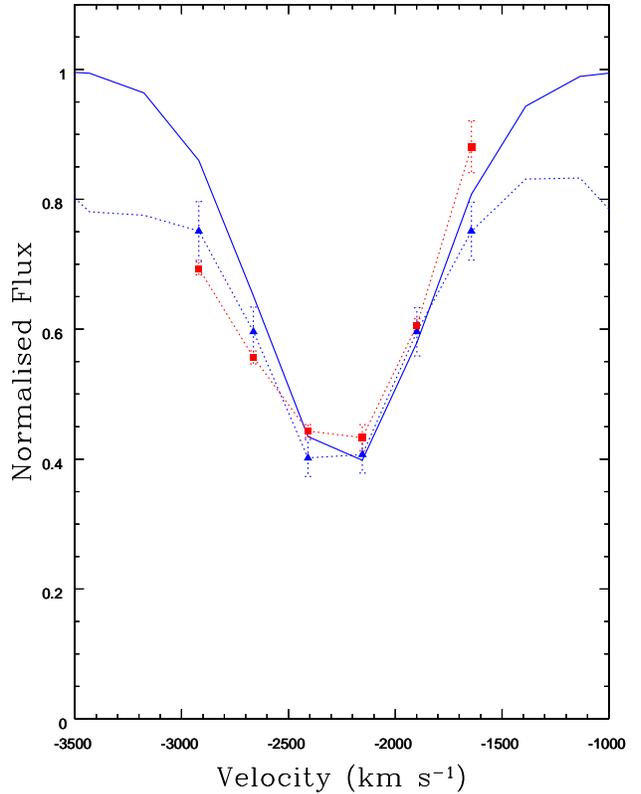}}
\caption{Inhomogeneous absorber model (red dotted line) of the blue doublet component for a power-law index a=3.3. The solid blue line is the blue component Gaussian model and the dotted blue line is the observed spectrum in velocity range of the blue component, both of which have been interpolated onto the red component velocity grid. Power-law model error bars are based on the errors of $a$. Blue triangles and red squares indicate the points on the observed spectrum and inhomogeneous absorber model respectively.}
\label{fig:inhom}
\end{figure}

\begin{figure}
\centering
\resizebox{\hsize}{!}{\includegraphics[angle=0,width=8cm]{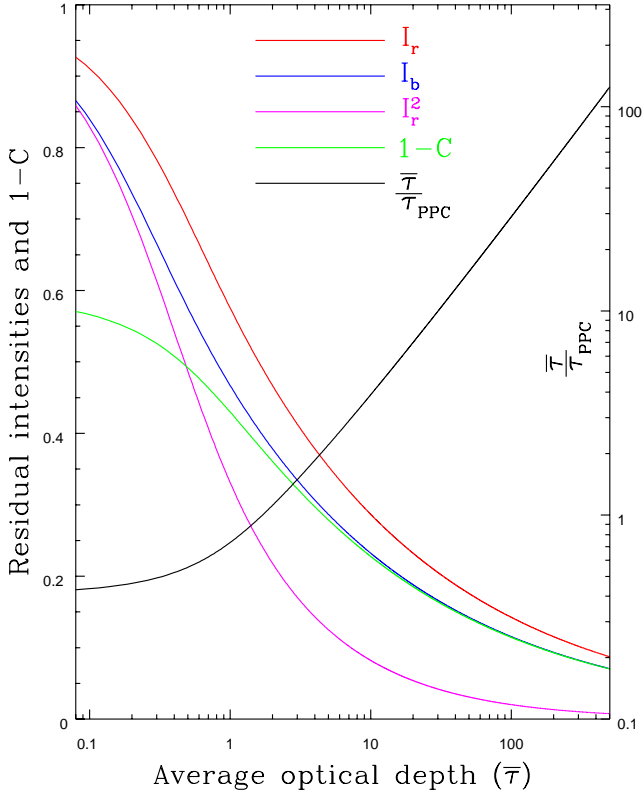}}
\caption{Behaviour of doublet components with respect to average optical depth for the power-law model adopted (a=3.3). The residual intensity of the blue component $I_{b}$ (blue line) is found by doubling the optical depth of the simulated red component $I_{r}$ (red line). The $I_{r}^{2}$ value (magenta line) indicates what the blue component would have been for a homogeneous absorber completely covering the emission source. The green line is (1-C) for a PPC model given the values of $I_{r}$ and $I_{b}$ at a specific $\bar{\tau}$. The ratio of $\frac{\bar{\tau}}{\tau_{\rm{PPC}}}$ indicated by the black line, where $\tau_{\rm{PPC}}$ is the optical depth of the PPC model predicting the ($1-C$) curve, shows the divergence of the power-law model from the PPC model.} 
\label{fig:inhomod}
\end{figure}

Similar to the procedure in Section 3.4, we first examine the plausibility of a power-law model describing the real behaviour of the absorption features by attempting to predict the profile of the blue component of absorber 1 in epoch 2 after interpolation of the blue absorption line profile into the velocity space of the red component. Given a value of $a$ to be tested, the process is carried out as follows: (i) Find the value of $\tau$$_{max}$ which gives the value of the residual red component flux $I_{r}$=$\int_{0}^{1}$$e^{-\tau_{max}x^{a}}$$dx$ at each velocity bin, (ii) Predict each $I_{b}$ value using $I_{b}=\int_{0}^{1}e^{-2\tau_{max}x^{a}}dx$, and (iii) Compare predicted $I_{b}$ values to the observed values and calculate the corresponding $\chi$$^{2}$ value over the entire velocity range. This procedure is repeated for all values of $a$, the value adopted being the one which minimises the $\chi$$^{2}$ calculation in step (iii). This provides a best value of a=3.3$^{+0.4}_{-0.1}$, the corresponding predicted blue profile is shown in Fig.~\ref{fig:inhom}. Finding $N_{ion}$ requires knowledge of the average optical depth $\bar{\tau}$, where $\bar{\tau}=\int_{0}^{1}\tau(x)dx$. As in Fig. 2 and Fig. 3 of \citet{arav05}, the dependence of the doublet component residual intensities on $\bar{\tau}$ as well as comparisons with complete homogeneous coverage and PPC models are shown in Fig.~\ref{fig:inhomod}.  

From Fig.~\ref{fig:inhom} it is clear that, like the PPC model, an inhomogeneous (power-law) absorber model is a good predictor of the behaviour of the blue component in this particular doublet. As in Section 3.4, this is extended to the Gaussian models with doublet components treated as single points with flux averaged over all velocity bins showing absorption. Interestingly, this leaves the upper bound on $a$ unconstrained for all absorption doublets except absorber 4 at epoch 1 (see Table~\ref{tab:inhabs}).

\begin{table}
\begin{center}
\caption{Power indices calculated by matching the predicted blue component of an inhomogeneous power-law to the Gaussian model equivalent ('unc' indicates 'unconstrained'). Several of the values have reached the upper limit of 10, necessarily leaving the upper bound unconstrained in these cases.}
\begin{tabular}{l lll}
\hline Absorber&\multicolumn{3}{c}{Power-law index}\\
 &Epoch 1&Epoch 2&Epoch 3\\
\hline 
1&10.0$^{+unc}_{-0.9}$&9.9$^{+unc}_{-3.3}$&10.0$^{+unc}_{-1.3}$\\
2&10.0$^{+unc}_{-1.2}$&7.1$^{+unc}_{-6.1}$&6.5$^{+unc}_{-3.6}$\\
3&10.0$^{+unc}_{-2.2}$&5.4$^{+unc}_{-2.9}$&1.0$^{+unc}_{-0.4}$\\
4&6.4$^{+2.6}_{-1.7}$&3.9$^{+unc}_{-2.2}$&4.6$^{+unc}_{-3.2}$\\
5&0.7$^{+unc}_{-0.4}$&8.3$^{+unc}_{-5.1}$&10.0$^{+unc}_{-6.9}$\\
\hline
\end{tabular}
\label{tab:inhabs}
\end{center}
\end{table}

Given that the best-fit power-law index of the Gaussian model for absorber 1 at epoch 2 is not consistent with the measured value of $a$ from the six velocity bins in the observed spectrum, this analysis is considered inconclusive. The constraints on the strongest doublet (absorber 1) suggest that this feature has a comparatively large value of $a$. At large values of $a$ the dependence of optical depth on position, $x$, approximates a step function, and thus cannot easily be distinguished from a PPC model.

\section{Photoionization Simulations}

\subsection{Cloudy Setup}

Modelling the column densities and ionization fractions of ions in quasar outflows by specifying the shape of the ionizing continuum and various gas (column) densities can provide insight into the physical location of the outflowing plasma and the changing nature of the relationship between the ionizing photon flux and ionization state, especially if ionization changes are the principal driver of absorption line variability. Photoionization models in this investigation are performed using {\sc cloudy} \citep{ferland98}, version c13.02.

The {\sc cloudy} code combines these input parameters with a model of the input spectral energy distribution (SED) to calculate the total ionizing photon flux. It models the input SED using input parameters for temperature ($T$) and power-law indices determining continuum behaviour of the form $F_{\nu}$~$\propto$~$\nu$$^{\alpha}$, including optical to near-UV spectral index ($\alpha_{uv}$) for optical wavelengths above 2500 \AA{}, UV to soft x-ray spectral index ($\alpha_{ox}$) between 2500 \AA{} and 2 keV, and x-ray spectral index ($\alpha_{x}$) from 2 keV to 100 keV. The $\alpha_{uv}$ value used is 0, which is within the range found by other studies, for example \citet{natali98} found an average value of $\alpha_{uv}$=$-0.33$$\pm$$0.59$. We adopt a value of $\alpha_{ox}=-1.98$ calculated using the method of \citet{wilkes94} in Equation~\ref{eqn:aox}

\begin{equation}
\label{eqn:aox}
\alpha_{ox}=-1.53-0.11\times{}log\left(\frac{L_{o}}{10^{30.5}}\right)
\end{equation}

\noindent where $L_{o}$ is the specific intensity at 2500 \AA{} in the quasar rest-frame in units of erg s$^{-1}$ Hz$^{-1}$. This is towards the more negative end of the range found in \citet{grupe10}, however tests with values of $\alpha_{ox}$ between a more typical value of $-1.6$ and the adopted value show no significant difference in Si~{\sc iv} and C~{\sc iv} column densities, so any discrepancies between our adopted value and the true value should not dramatically alter our results. According to \citet{zdziarski96} the value of $\alpha_{x}$ is typically between $-0.8$ and $-1.0$, so a value of $-1.0$ is adopted here for simplicity. It is possible to estimate $T$ from the ratio of the bolometric to Eddington luminosity and the mass of the black hole using the method of \citet{bonning07}:

\begin{equation}
\label{eqn:tmax}
T_{max}=10^{5.56}M^{-\frac{1}{4}}_{8}\frac{L_{bol}}{L_{Edd}}
\end{equation}

\noindent where $T_{max}$ is the maximum accretion disk temperature, $M_{8}$ is the black hole mass in units of $10^{8}$$M_{\odot}$, $L_{bol}$ is the bolometric luminosity and $L_{Edd}$ is the Eddington luminosity. The value of $L_{Edd}$ can be found from the black hole mass using $L_{Edd}=10^{46.1}M_{8}$. Using estimates of $M_{8}$ and $L_{bol}$ from \citet{shen11} gives a value of $T_{max}$$\sim$240\,000 K, with the peak flux of the big blue bump occurring at $2.1$$\times$$10^{16}$ Hz ($\sim$140 \AA{} or 87 eV).

\begin{figure}
\centering
\resizebox{\hsize}{!}{\includegraphics[angle=0,width=8cm]{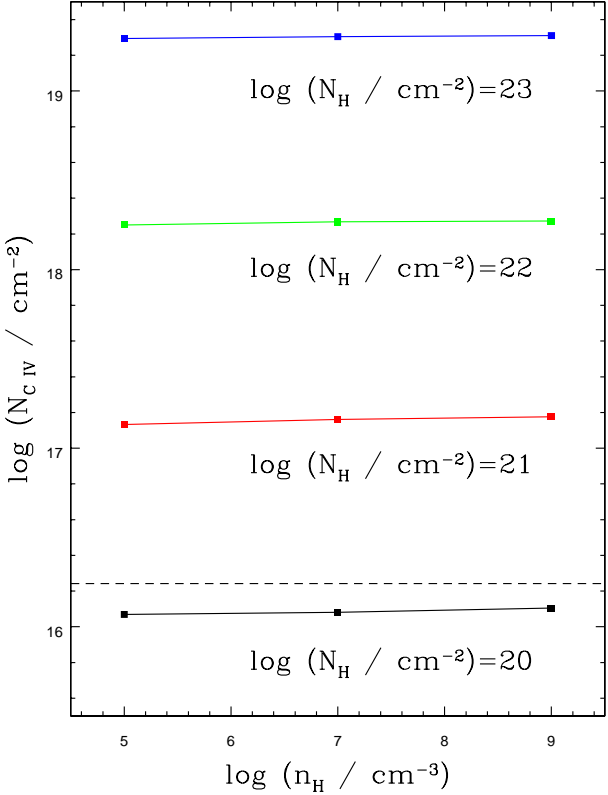}}
\caption{Predicted C~{\sc iv} column densities as a function of hydrogen number density for given hydrogen column densities. Solid squares indicate the {\sc cloudy} input hydrogen densities, which are linked by solid lines of constant hydrogen column density. The horizontal dashed line gives the minimum C~{\sc iv} column density predicted by the SDSS observation, which shows the strongest absorption of all epochs. It is clear that log~($N_{H}$ / cm$^{-2}$)$\leq$20 is ruled out by this limit.}
\label{fig:c4peaks}
\end{figure}

A grid of models containing column density data for Si~{\sc iv} and C~{\sc iv} is generated using the input continuum, with grid-points at hydrogen column densities log~($N_{H}$ / cm$^{-2}$)=21, 22 and 23, hydrogen number densities log~($n_{H}$ / cm$^{-3}$)=5, 7 and 9 and log $U$ values spanning $-$5.0 to 3.0 in intervals of 0.2 dex. These hydrogen number densities span typical narrow line region to broad line region (BLR) densities, the maximum density being close to the critical density of C~{\sc iii]}. The density is unlikely to be higher than this as this would suggest the absorber originates from within the BLR, conflicting with the fact that individual absorption components are relatively narrow in comparison to the broad emission lines. This is also apparent from the need for the deepest troughs to be absorbing both the BEL and the continuum. The ionization parameter range encompasses all reasonable values given the existence of the absorbers, while the hydrogen column density range was chosen to span a range above the minimum ionic column density for C~{\sc iv}, $N_{\rm C~{IV}}$, and below a density at which Thomson scattering becomes significant. This rules out column densities of log~($N_{H}$ / cm$^{-2}$)$\leq$20 as these generate $N_{\rm C~{IV}}$ values below the minimum value calculated from direct integration (Fig.~\ref{fig:c4peaks}).

\subsection{Application of Cloudy Models to Estimated Column Densities}

The {\sc cloudy} models can be applied to results from the previous sections in this paper to provide parameters governing the properties of the quasar continuum source which affect the outflowing gas. Initially it is necessary to find the correct normalisation of the continuum SED described in Section 4.1, which can be used to find the total ionizing luminosity $L_{ion}$ and ionizing photon emission rate $Q(H)$ and hence the distance from the continuum source to the inner edge of the outflowing gas (facing the source) and the mass outflow rate. The correct ionising SED is found by scaling the output to the rest-frame corrected flux of the observed SDSS spectrum. Output parameters from the simulation include the total ionizing photon flux $\phi$$(H)$ which can be used to find the true values of the ionising luminosity, $L_{ion}$, the rate of emission of Lyman continuum photons, $Q(H)$ and the ionising source distance to the absorbing gas, $R$. 

The calculated value of $R$ will vary depending on the grid parameters ($U$,$N_{H}$ and $n_{H}$); however since the same SED is used in each case, values of ionizing luminosity and ionizing photon production rate are found which are applied universally; they are $L_{ion}$=4.17$\times$10$^{46}$ erg s$^{-1}$and $Q(H)$=9.80$\times$10$^{56}$ s$^{-1}$. The elemental abundances used by {\sc cloudy} are a solar composition derived from \citet{grevesse98}, \citet{holweger01}, and \citet{allendeprieto01,allendeprieto02}.

In order to apply the grid of models calculated in Section 4.1, limits for the column densities of Si~{\sc iv} and C~{\sc iv} must be found. Since there are no resolvable doublets available for C~{\sc iv}, only the lower limits found from direct integration (see Section 3.2) can be used for this ion. For Si~{\sc iv}, the strength of the doublets at epoch 1 means total column densities can be estimated using the PPC model for all 5 absorbers as listed in Table~\ref{tab:ppcabs}. For epochs 2 and 3, PPC can be used for absorbers 1 and 2 which are the predominant contributors to the total line of sight column density at these epochs, plus a contribution from absorbers 3 to 5 using the peak optical depth (POD) method outlined in Section 3.3 modified to take account of partial coverage. Here we assume that the covering fractions of absorbers 3 to 5 at epochs 2 and 3 are the same as those at epoch 1. The true optical depth can be found once the covering fraction is known. Multiplying the original POD-derived $N_{\rm Si~{IV}}$ by the covering fraction and the ratio of the integrated real optical depth over the integrated apparent optical depth spanning the velocity range of the red component of the doublet, gives an estimate of the true column density. The $N_{\rm Si~{IV}}$ and their upper and lower limits are shown in Table~\ref{tab:si4ntots}. The total column density lower limits are assumed to be the totals derived from the POD method (listed in Section 3.3). Upper limits on the value of $N_{\rm Si~{IV}}$ at epoch 2 are comparatively large due to the large uncertainty in the covering fraction of absorber 2, however the best value lies towards the lower end of this interval.

\begin{table}
\begin{center}
\caption{Total column densities and their limits for Si~{\sc iv} at each epoch.}
\begin{tabular}{l l l l}
\hline Epoch&$N_{\rm Si~{IV}}$&$N_{\rm Si~{IV}}$ lower limit&$N_{\rm Si~{IV}}$ upper limit\\
 &($\times$10$^{14}$ cm$^{-2}$)&($\times$10$^{14}$ cm$^{-2}$)&($\times$10$^{14}$ cm$^{-2}$)\\
\hline
1&59.8&33.5&94.2\\
2&9.18&7.79&27.5\\
3&18.3&14.7&27.9\\
\hline 
\end{tabular}
\label{tab:si4ntots}
\end{center}
\end{table}

\begin{figure}
\centering
\resizebox{\hsize}{!}{\includegraphics[angle=0,width=8cm]{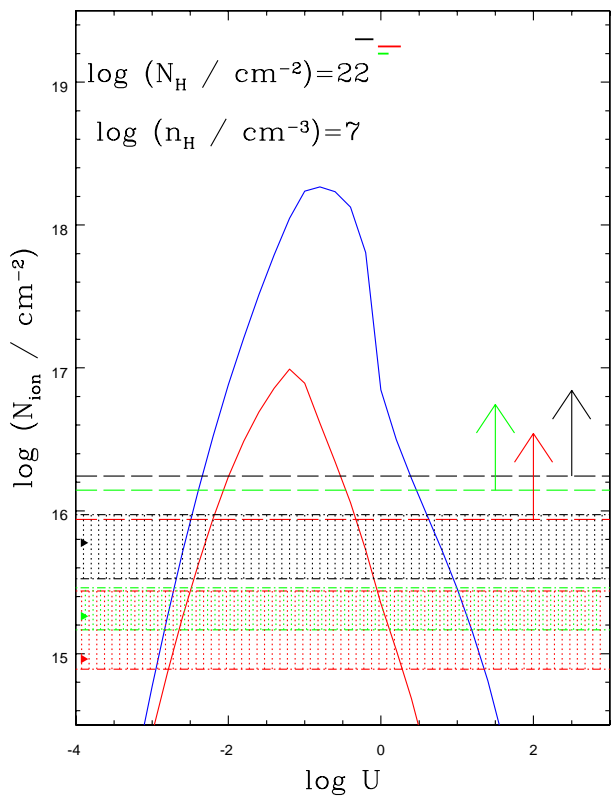}}
\caption{Predicted Si~{\sc iv} (solid red lines) and C~{\sc iv} (solid blue lines) column densities as a function of ionization parameter at log~($N_{H}$ / cm$^{-2}$)=22 and log~($n_{H}$ / cm$^{-3}$)=7. Black, red and green dashed lines with arrows indicate the lower limits of the C~{\sc iv} column density for epochs 1, 2 and 3 respectively. Using the same colour scheme, the solid triangles represent the best values of the Si~{\sc iv} column densities, while the dashed-dotted lines show the confidence intervals on these values with the area enclosed being shaded by vertical dotted lines. The thick solid lines at the top show the corresponding log~$U$ limits.}
\label{fig:cdnplot}
\end{figure}

\begin{figure}
\centering
\resizebox{\hsize}{!}{\includegraphics[angle=0,width=8cm]{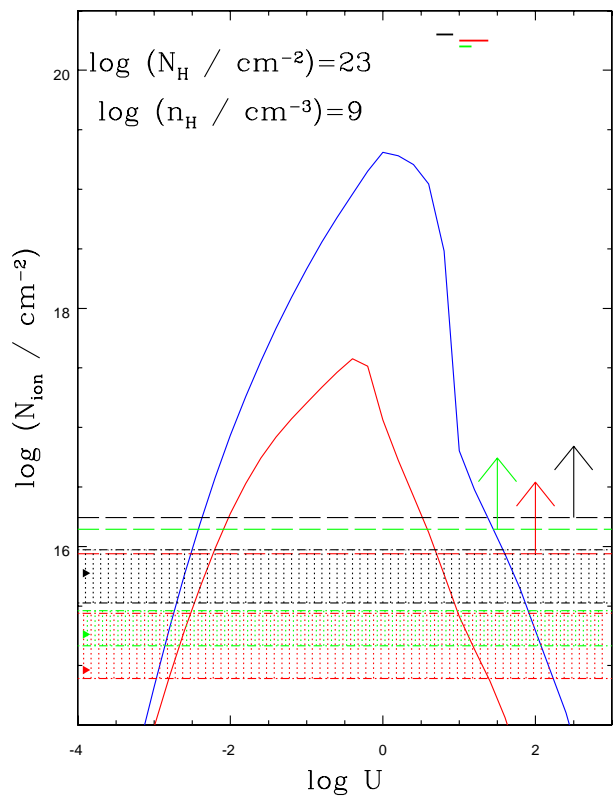}}
\caption{Predicted Si~{\sc iv} (solid red lines) and C~{\sc iv} (solid blue lines) column densities as a function of ionization parameter at log~($N_{H}$ / cm$^{-2}$)=23 and log~($n_{H}$ / cm$^{-3}$)=9. Black, red and green dashed lines with arrows indicate the lower limits of the C~{\sc iv} column density for epochs 1, 2 and 3 respectively. Using the same colour scheme, the solid triangles represent the best values of the Si~{\sc iv} column densities, while the dashed-dotted lines show the confidence intervals on these values with the area enclosed being shaded by vertical dotted lines. The thick solid lines at the top show the corresponding log~$U$ limits.}
\label{fig:cdnplot2}
\end{figure}

Using the {\sc cloudy} output at each combination of log~$N_{H}$ and log~$n_{H}$, plots can be made of predicted ionic column densities across the entire range of ionization parameter $U$. The span of log~$U$ covers the peak of both $N_{ion}$ for all combinations, resulting in cases where there are two possible regions in which the ionic column density values can be located. Given that only lower limits of the C~{\sc iv} column density can be found, the column densities for this ion are estimated by locating the positions in log~$U$ space of allowed $N_{\rm Si~{IV}}$ and using the equivalent values for C~{\sc iv}. If it is assumed that ionizing continuum changes are the main factor driving variability, it is possible to identify the range of log~$U$ which satisfies the ionic column density limits at each epoch. Although there are nine possible combinations of log~$N_{H}$ and log~$n_{H}$, only 2 are shown (Figures~\ref{fig:cdnplot} and~\ref{fig:cdnplot2}). It is apparent from these Figures that the lower region of allowed log~$U$ values (where the gradient of log~$N_{ion}$ vs. log~$U$ is positive) is invalid as the allowed range of $N_{\rm Si~{IV}}$ for epoch 3 does not permit any $N_{\rm C~{IV}}$ values above the lower limit. This was found to be true for all 9 density/column density combinations, giving one continuous span of allowed log~$U$ values in each case.  

\subsection{Outflow properties derived from models}

The estimation of mass outflow rates ($\dot M_{out}$) is based on the method used in \citet{crenshaw09} which relates $\dot M_{out}$ to other parameters using the following equation

\begin{equation}
\label{eqn:mout}
\dot M_{out}=8\pi{}RN_{H}\mu{}m_{p}C_{q}|v|\, (M_{\odot} yr^{-1})
\end{equation}

\noindent where $\mu$ is the mean atomic mass per proton ($\mu$$\sim$1), $m_{p}$ is the proton mass, $C_{q}$ is the covering fraction of the outflow as seen from the quasar and v is the mean velocity of the outflow. The value of $C_{q}$ is assumed to be $\sim$0.1 given the fraction of quasars showing broad absorption lines. It is not obvious what the mean outflow velocity of the predominantely hydrogen outflow is, so we have adopted the centroid velocity of the C~{\sc iv} outflow, v$\sim$$-$7600 km s$^{-1}$.

In Table~\ref{tab:outputs} information on mass outflow rate, distance to the inner face of the outflowing gas from the ionising continuum source and C~{\sc iv} column density is provided at the values of $U$ allowed by the limits on $N_{\rm Si~{IV}}$ and lower limits on $N_{\rm C~{IV}}$ at each grid point. This gives an overview of the allowed range in physical state of the outflow, as well as possible effects a changing ionization parameter could exert. Linear interpolation between the 0.2 dex interval-separated log~$U$ values is used to estimate values at intervals of 0.02 dex. 

\begin{table*}
\begin{center}
\caption{Table of parameters derived from {\sc cloudy} simulations for various hydrogen densities and column densities.}
\begin{tabular}{l l l l l l l l}
\hline log~$N_{H}$&log~$n_{H}$&log~$U$&log~$U$ range&log~$N_{\rm C~{IV}}$&log~$N_{\rm C~{IV}}$ range&log~$R$&Outflow Rate\\
log(cm$^{-2}$)&log(cm$^{-3}$)&&&log(cm$^{-2}$)&log(cm$^{-2}$)&log(cm)&M$_{\odot}$/year\\
\hline\hline
Epoch 1\\
\hline
\phantom{$^{\dagger}$}21&5&-1.20&-1.36$\leq$log~$U$$\leq$-1.06&17.09&16.91$\leq$log~$N_{\rm C~{IV}}$$\leq$17.12&20.8$\pm$0.1&36.1$^{+9.4}_{-7.4}$\\
\phantom{$^{\dagger}$}21&7&-1.24&-1.38$\leq$log~$U$$\leq$-1.06&17.10&16.88$\leq$log~$N_{\rm C~{IV}}$$\leq$17.15&19.8$\pm$0.1&3.61$^{+0.94}_{-0.74}$\\
\phantom{$^{\dagger}$}21&9&-1.18&-1.34$\leq$log~$U$$\leq$-1.02&17.07&16.85$\leq$log~$N_{\rm C~{IV}}$$\leq$17.15&18.8$\pm$0.1&0.361$^{+0.093}_{-0.074}$\\
$^{\dagger}$22&5&-0.20&-0.30$\leq$log~$U$$\leq$-0.08&17.90&17.32$\leq$log~$N_{\rm C~{IV}}$$\leq$18.03&20.3$\pm$0.1&114$^{+30}_{-23}$$^{*}$\\
\phantom{$^{\dagger}$}22&7&-0.24&-0.34$\leq$log~$U$$\leq$-0.10&17.87&17.33$\leq$log~$N_{\rm C~{IV}}$$\leq$18.03&19.3$\pm$0.1&11.4$^{+3.0}_{-2.3}$\\
\phantom{$^{\dagger}$}22&9&-0.20&-0.30$\leq$log~$U$$\leq$-0.08&17.76&17.22$\leq$log~$N_{\rm C~{IV}}$$\leq$17.94&18.3$\pm$0.1&1.14$^{+0.30}_{-0.23}$\\
$^{\dagger}$23&5&0.80&0.68$\leq$log~$U$$\leq$0.92&18.79&17.66$\leq$log~$N_{\rm C~{IV}}$$\leq$18.98&19.8$\pm$0.1&361$^{+94}_{-74}$$^{*}$\\
\phantom{$^{\dagger}$}23&7&0.76&0.66$\leq$log~$U$$\leq$0.90&18.72&17.73$\leq$log~$N_{\rm C~{IV}}$$\leq$18.94&18.8$\pm$0.1&36.1$^{+9.4}_{-7.4}$\\
\phantom{$^{\dagger}$}23&9&0.80&0.70$\leq$log~$U$$\leq$0.92&18.48&17.48$\leq$log~$N_{\rm C~{IV}}$$\leq$18.76&17.8$\pm$0.1&3.61$^{+0.94}_{-0.74}$\\
\hline
Epoch 2\\
\hline
\phantom{$^{\dagger}$}21&5&-0.64&-0.98$\leq$log~$U$$\leq$-0.62&16.37&16.31$\leq$log~$N_{\rm C~{IV}}$$\leq$16.81&20.5$^{+0.2}_{-0.1}$&18.1$^{+10.6}_{-3.7}$\\
\phantom{$^{\dagger}$}21&7&-0.66&-1.00$\leq$log~$U$$\leq$-0.62&16.34&16.29$\leq$log~$N_{\rm C~{IV}}$$\leq$16.80&19.5$^{+0.2}_{-0.1}$&1.81$^{+1.05}_{-0.37}$\\
\phantom{$^{\dagger}$}21&9&-0.62&-0.94$\leq$log~$U$$\leq$-0.58&16.33&16.28$\leq$log~$N_{\rm C~{IV}}$$\leq$16.74&18.5$^{+0.2}_{-0.1}$&0.181$^{+0.106}_{-0.037}$\\
$^{\dagger}$22&5&0.26&-0.02$\leq$log~$U$$\leq$0.30&16.46&16.40$\leq$log~$N_{\rm C~{IV}}$$\leq$17.03&20.0$^{+0.2}_{-0.1}$&57.4$^{+33.3}_{-11.9}$\\
\phantom{$^{\dagger}$}22&7&0.24&-0.04$\leq$log~$U$$\leq$0.26&16.44&16.42$\leq$log~$N_{\rm C~{IV}}$$\leq$17.04&19.1$^{+0.2}_{-0.1}$&7.20$^{+4.20}_{-1.48}$\\
\phantom{$^{\dagger}$}22&9&0.28&-0.02$\leq$log~$U$$\leq$0.30&16.42&16.39$\leq$log~$N_{\rm C~{IV}}$$\leq$16.95&18.0$^{+0.2}_{-0.1}$&0.574$^{+0.333}_{-0.119}$\\
$^{\dagger}$23&5&1.32&1.00$\leq$log~$U$$\leq$1.36&16.33&16.27$\leq$log~$N_{\rm C~{IV}}$$\leq$16.90&19.5$^{+0.2}_{-0.1}$&181$^{+106}_{-37}$$^{*}$\\
\phantom{$^{\dagger}$}23&7&1.30&0.98$\leq$log~$U$$\leq$1.34&16.32&16.27$\leq$log~$N_{\rm C~{IV}}$$\leq$17.00&18.5$^{+0.3}_{-0.1}$&18.1$^{+18.0}_{-3.7}$\\
\phantom{$^{\dagger}$}23&9&1.34&1.00$\leq$log~$U$$\leq$1.38&16.29&16.23$\leq$log~$N_{\rm C~{IV}}$$\leq$16.81&17.5$^{+0.2}_{-0.1}$&1.81$^{+1.06}_{-0.37}$\\
\hline
Epoch 3\\
\hline
\phantom{$^{\dagger}$}21&5&-0.86&-0.98$\leq$log~$U$$\leq$-0.82&16.63&16.57$\leq$log~$N_{\rm C~{IV}}$$\leq$16.81&20.6$\pm$0.1&22.8$^{+5.9}_{-4.7}$\\
\phantom{$^{\dagger}$}21&7&-0.88&-1.00$\leq$log~$U$$\leq$-0.82&16.63&16.54$\leq$log~$N_{\rm C~{IV}}$$\leq$16.80&19.6$\pm$0.1&2.28$^{+0.59}_{-0.47}$\\
\phantom{$^{\dagger}$}21&9&-0.84&-0.96$\leq$log~$U$$\leq$-0.78&16.60&16.52$\leq$log~$N_{\rm C~{IV}}$$\leq$16.77&18.6$\pm$0.1&0.228$^{+0.059}_{-0.047}$\\
$^{\dagger}$22&5&0.08&-0.02$\leq$log~$U$$\leq$0.12&16.77&16.70$\leq$log~$N_{\rm C~{IV}}$$\leq$17.03&20.1$\pm$0.1&72.0$^{+18.7}_{-14.8}$\\
\phantom{$^{\dagger}$}22&7&0.06&-0.04$\leq$log~$U$$\leq$0.10&16.74&16.67$\leq$log~$N_{\rm C~{IV}}$$\leq$17.04&19.2$\pm$0.1&9.05$^{+2.35}_{-1.85}$\\
\phantom{$^{\dagger}$}22&9&0.08&-0.02$\leq$log~$U$$\leq$0.14&16.72&16.62$\leq$log~$N_{\rm C~{IV}}$$\leq$16.95&18.1$\pm$0.1&0.720$^{+0.187}_{-0.148}$\\
$^{\dagger}$23&5&1.12&1.00$\leq$log~$U$$\leq$1.16&16.66&16.59$\leq$log~$N_{\rm C~{IV}}$$\leq$16.90&19.6$\pm$0.1&228$^{+59}_{-47}$$^{*}$\\
\phantom{$^{\dagger}$}23&7&1.08&0.98$\leq$log~$U$$\leq$1.14&16.68&16.57$\leq$log~$N_{\rm C~{IV}}$$\leq$17.00&18.6$\pm$0.1&22.8$^{+5.9}_{-4.7}$\\
\phantom{$^{\dagger}$}23&9&1.12&1.00$\leq$log~$U$$\leq$1.16&16.61&16.54$\leq$log~$N_{\rm C~{IV}}$$\leq$16.81&17.6$\pm$0.1&2.28$^{+0.58}_{-0.47}$\\
\hline 
\end{tabular}
\label{tab:outputs}
\end{center}
$^{*}$ These outflow rates are much greater than the accretion rate and are therefore ruled out.\\
$^{\dagger}$ These $N_{H}$,$n_{H}$ grid-points are ruled out due to the large outflow rates indicated by $^{*}$.\\
\end{table*}

\begin{figure}
\centering
\resizebox{\hsize}{!}{\includegraphics[angle=0,width=8cm]{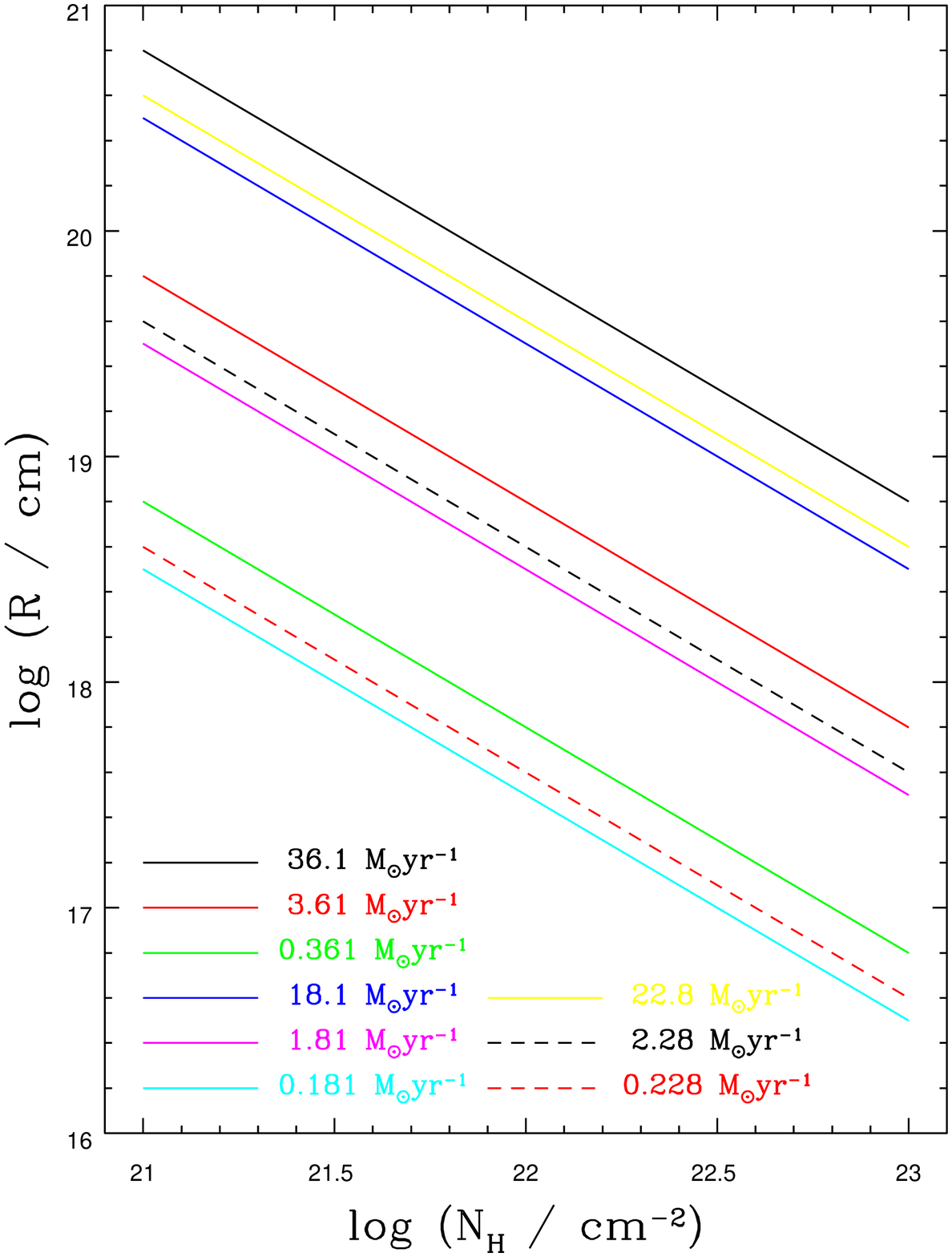}}
\caption{Dependence of log~$R$ on log~$N_{H}$ for constant mass outflow rates. Values are from Table~\ref{tab:outputs}.}
\label{fig:constmout}
\end{figure}

As expected, the range of log~$U$ allowed is strongly dependent on $N_{H}$ rather than $n_{H}$, leaving the value of the hydrogen number density uncertain, with the caveat that it probably does not exceed the BLR density. If ionizing continuum flux variations are driving the variability, then given the values of log~$U$ and their possible ranges indicated by the third and fourth columns of Table~\ref{tab:outputs} it is possible to estimate how the ionization parameter is changing across the time scales sampled. Between epochs 1 and 2, log~$U$ will be changing by increments of between 0.06 dex and 0.76 dex with best estimates of $\sim$0.5 to $\sim$0.6 dex. For the epoch 2 to epoch 3 interval, log $U$ changes by increments of between 0 and 0.38 dex, with $\sim$$-0.2$ dex being the best estimate. This would require a similar change in the ionizing flux reaching the outflowing gas. Changes of similar scale to these best estimates for the change in log $U$ have been observed in the extreme ultraviolet (EUV) of at least one other quasar of similar redshift over a comparable rest-frame timescale e.g. in \citet{reimers08}, where the flux at a rest-frame wavelength of 335 \AA{} changed by a factor of $\sim$3 in a rest-frame timescale of $\sim$0.65 years. Since the EUV is the main contributor to outflow ionization, this indicates our results are plausible.

As indicated in Column 7 of Table~\ref{tab:outputs}, the log~$R$ values do not change significantly between epochs for given combinations of log~$N_{H}$ and log~$n_{H}$. The dependence of log$R$ on log$N_{H}$ is indicated in Fig.~{\ref{fig:constmout}. It is appropriate to consider how the predicted absorber distances compare to the size of the BLR. Using the method of \citet{kaspi05}, the BLR radius $R_{BLR}$ can be estimated using $R_{BLR}$=AL$^{B}$$\times$10 lt-days, where $L$ is $\lambda$$L_{\lambda}$(1450 \AA{})/10$^{44}$(erg s$^{-1}$) and $A$ and $B$ are constants, with a set of values found using the FITEXY method and the mean Balmer line's time lag in \citet{kaspi05} to be $A$=2.12 and $B$=0.496. This set of values is used as the index $B$ is consistent with subsequent studies, e.g. \citet{bentz06} which found it to be $\sim$0.5 This is consistent with the fact that $U$$\propto$$Q_{H}$/$R^{2}$, leading to $R$$\propto$$L^{0.5}$. The procedure provides a value of $R_{BLR}$$\sim$7.6$\times$$10^{17}$ cm (a few hundred light-days). Comparing this with the values found for $R$ in Table~\ref{tab:outputs} suggests that the outflowing gas is outside the BLR in 8 out of 9 cases, the exception being the grid-point log~($N_{H}$ / cm$^{-2}$)=23, log~($n_{H}$ / cm$^{-3}$)=9.

The mass outflow rates calculated range from tenths of a solar mass to hundreds of solar masses per year. The estimated mass outflow rates, which depend upon the product $RN_{H}$, are in the range indicated in Table~\ref{tab:outputs}. Given the uncertainties in the mass outflow rate, they do not change significantly between epochs. It is possible to calculate the accretion rate at Eddington luminosity in this object and hence compare it to the actual accretion rate and the mass outflow rate. This requires an estimation of the black hole mass, which can be calculated using the following

\begin{equation}
\label{eqn:bhmass}
M=\frac{fv^{2}R_{BLR}}{G}
\end{equation}

\noindent where $v$ is the orbital velocity of the BLR gas, $G$ is the gravitational constant and $f$ is the virial coefficient which has been estimated to be $\sim$5.25 \citep{woo10}. The value of $v$ is estimated as $v$$\sim$4100 km s$^{-1}$ using the width of the C~{\sc iv} broad emission line. This gives a value of $M_{BH}$$\sim$5.1$\times$10$^{9}$ $M_{\odot}$. This was checked for consistency by comparing with the value obtained using Equation 3 in \citet{vestergaard04}, which provides a very similar value of $\sim$5.3$\times$$10^{9}$ $M_{\odot}$. The Eddington luminosity can then be found using the relation $L_{Edd}$=10$^{46.1}$$M_{8}$ erg s$^{-1}$, where $M_{8}$ is black hole mass in units of 10$^{8}$$M_{\odot}$, giving $L_{Edd}$=6.4$\times$10$^{47}$ erg s$^{-1}$. The Eddington accretion rate $\dot M_{Edd}$=$L_{Edd}$/$\eta$c$^{2}$ implies $\dot M_{Edd}$$\sim$113 $M_{\odot}$yr$^{-1}$. 

The actual accretion rate can be calculated from the bolometric luminosity, which can be found using the luminosity at 1450\AA{} by applying $L_{bol}$$\sim$4.36$\lambda$$L_{\lambda}$ (at 1450\AA{}) \citep{warner04}. For a bolometric luminosity of $L_{bol}$$\sim$9.5$\times$10$^{46}$ erg s$^{-1}$, the resultant value of the accretion rate in terms of that at Eddington is $\sim$0.15 $\dot M_{Edd}$ ($\sim$17.0 $M_{\odot}$yr$^{-1}$), meaning that this object's accretion rate is a substantial portion of the value at the Eddington limit. This allows the ruling out of two combinations of hydrogen column density and hydrogen number density (log~($N_{H}$ / cm$^{-2}$)=22, log~($n_{H}$ / cm$^{-3}$)=5 and log~($N_{H}$ / cm$^{-2}$)=23, log~($n_{H}$ / cm$^{-3}$)=5) due to their excessively high outflow rates compared to the accretion rate, as indicated in Table~\ref{tab:outputs}. Comparing the remaining allowed values suggests the mass outflow rate is approximately between 1 per cent and 200 per cent of the mass accretion rate, corresponding to a kinetic luminosity range of $\sim$3.3$\times$10$^{42}$ to 6.6$\times$10$^{44}$ erg s$^{-1}$. The upper limit of this range gives a kinetic energy to luminosity ratio of $\sim$0.7 per cent, high enough to have a significant impact on the evolution of the surrounding galaxy \citep{hopkins10}, however definite conclusions regarding the outflow's importance in this regard are impossible given the weakness of the constraint. 

\section{Discussion}

\subsection{Are changes in the outflow geometry responsible for variability?}

Changes in covering fraction within the context of the PPC model are one possibility for the dramatic variability seen in this quasar's absorption lines. From Section 3.4 (Fig.~\ref{fig:intp}) it is clear that the deepest parts of the blue absorber 1 trough in epoch 2 are accurately described by a saturated absorber whose depth is determined by the fraction covering the emission source. However, Section 3.5 (Fig.~\ref{fig:inhom}) indicates that an inhomogeneous absorber modelled over the same velocity range is also an excellent predictor of the same trough. Further insight can be gained from the Gaussian model profiles, specifically absorbers 1 and 2 which are well-defined and hence can be studied using both PPC and inhomogeneous models. From Table~\ref{tab:inhabs} it appears that these features are not well described by an inhomogeneous absorber, since, given the maximum power-law index of 10, many of the calculated values are unconstrained. Large power indices produce line-of-sight geometries which resemble a step-function, which is equivalent to a PPC absorber. The difference between the power-law index which matches the deepest part of Absorber 1 in epoch 2 ($a$=3.3$^{+0.4}_{-0.1}$) and the value in Table~\ref{tab:inhabs} is also much greater than the uncertainty. This leads to the conclusion that an inhomogeneous absorber is probably not the predominant influence on absorption line profile shape.

If it is assumed that absorption takes place within the context of PPC, it is possible to determine how the covering fraction has changed between epochs using the Gaussian profile models. From Table~\ref{tab:ppcabs} is is clear that there is no significant change in $C$ for absorber 1 between epochs (changes being within 1$\sigma$ for $\sigma$$\leq$0.042). There is also no discernible change in the covering fraction of absorber 2. Conclusions regarding covering fraction variations in absorber 2 are weak for two reasons. First, the epoch 2 doublet is entirely unsaturated and therefore the covering fraction estimation is completely unconstrained. Secondly, when comparing the epoch 1 value ($C$=0.187$\pm$0.017) with the epoch 3 value ($C$=0.275$\pm$0.165), the large uncertainty in the latter allows a significant range which prevents any strong conclusion regarding the presence of covering fraction variations. Given the lack of significant variability in either the strongly constrained absorber 1 or the weakly constrained absorber 2, we conclude that there is no evidence for covering fraction being the dominant mechanism behind the variability.  

\subsection{Evidence for ionization changes driving variability}

Ionization fraction changes within the outflowing gas can be generated either by intrinsic changes in the ionizing radiation emitted from the central ionizing continuum source or shielding of this radiation by gas in the region between the accretion disk and the BAL gas. There are several indicators that suggest ionization fraction changes are implicated in the absorption line variability in this object, the strongest being coordinated changes seen across large velocity separations. A similar result was noted in \citet{trevese13} regarding the quasar APM 08279+5255, where they observed coordinated variability in two Gaussian-modelled components of a C~{\sc iv} BAL separated by $\sim$5000 km s$^{-1}$. Large velocity separations between absorption troughs imply that the absorption arises in physically distinct regions, hence making covering fraction changes an unlikely explanation due to the level of coordination required across such large separations (e.g. \citet{filizak13}). 

This phenomenon is especially evident in the C~{\sc iv} absorption troughs (see Fig.~\ref{fig:ionspec}). The visibly saturated absorption trough centred at $\sim$$-$4000 km s$^{-1}$ undergoes weakening between epochs 1 and 2 before strengthening again in epoch 3, which is a pattern also seen in the features at $\sim$$-$9500 km s$^{-1}$ and $\sim$$-$18 500 km s$^{-1}$. Since there is a lot of blending of the Si~{\sc iv} components, the same effect is not as obvious in this ion. However, it is instructive to look at the behaviour of the modelled Gaussian Si~{\sc iv} doublets in Fig.~\ref{fig:si4comps}, where the same effect is apparent.

No part of the power-law continuum contributing to this spectrum is observable short-ward of Ly-$\alpha$, however it is reasonable to expect that it extends in to the EUV portion of the spectrum  \citep{kriss99}, which is largely responsible for the flux which controls the ionization fractions of the plasma in the accretion disk's environment. Therefore variability in the observed continuum could indicate variability in the ionizing radiation from the accretion disk. There is evidence for an excess in the blue part of the spectrum during epoch 2 (first WHT observation) when compared to the prior and subsequent observations, as indicated in Figs~\ref{fig:spcont} and~\ref{fig:spcontn}, which could indicate an increase in the ionizing continuum flux at this epoch. At the 3rd epoch the blue excess appears to disappear, with the flux in this range appearing to follow the continuum from epoch 1 once again.

Evidence from the {\sc cloudy} simulations also supports an ionizing continuum change driving the variability. Given the decrease in the Si~{\sc iv} column density between epochs 1 and 2, an increase in ionization parameter must occur between these observations (See Figures \ref{fig:cdnplot} and \ref{fig:cdnplot2}), with an estimated log $U$ change of $\sim$0.5 to $\sim$0.6 dex. An increase in log $U$ agrees with the observed steepening of the continuum at the blue end of the spectrum between epochs 1 and 2, assuming the continuum extends to ionizing energies. There is a large overlap in predicted Si~{\sc iv} column densities between epochs 2 and 3, however the epoch 3 range lies at the higher end of the epoch 2 range, with the best epoch 3 value being 0.3 dex above the best epoch 2 value. A log $U$ change of $-$0.2 dex is needed to drive this increase in Si~{\sc iv} column density between epochs 2 and 3. A decrease in ionization parameter between these two epochs is supported by the apparent disappearance of the aforementioned blue excess in the continuum.

\begin{figure}
\centering
\resizebox{\hsize}{!}{\includegraphics[angle=0,width=8cm]{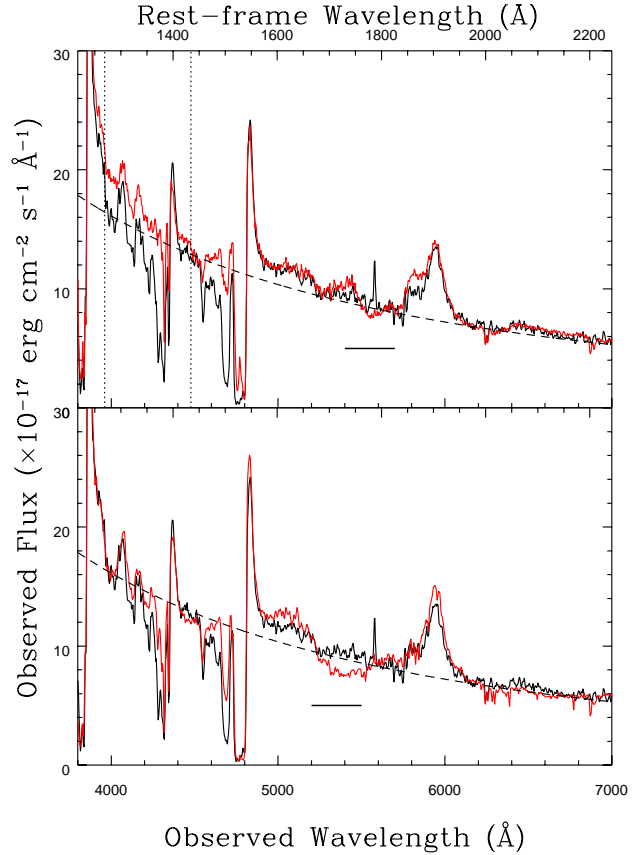}}
\caption{Upper panel: Epoch 1 observation (black line) and epoch 2 observation (red line) with power-law continuum (dashed black line) fitted to the epoch 1 spectrum. Vertical dotted lines indicate the boundaries of where the blue excess is apparent. The thick black horizontal line indicates the dichroic overlap region in the WHT observation. Lower panel: Same as upper panel but with red line indicating the epoch 3 observation.}
\label{fig:spcont}
\end{figure}

\begin{figure}
\centering
\resizebox{\hsize}{!}{\includegraphics[angle=0,width=8cm]{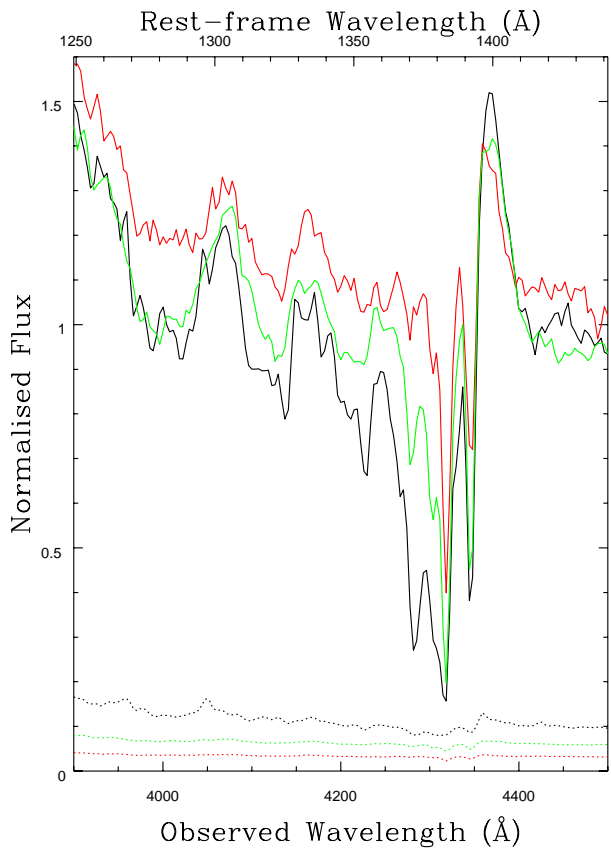}}
\caption{Epoch 1 observation (black), epoch 2 observation (red) and epoch 3 observation (green), normalised to the epoch 1 continuum shown in Figure~\ref{fig:spcont} over the blue excess range, with dotted lines representing the errors. The epoch 2 spectrum appears to lie almost everywhere above the epoch 1 continuum.}
\label{fig:spcontn}
\end{figure}

\subsection{Fourth Observation}

The quasar SDSS J1138+3517 was observed again by the WHT in May 2014 and is included in Fig.~\ref{fig:ionspec}. In contrast to the earlier WHT observations, the Si~{\sc iv} components are largely unchanged from the previous epoch's spectrum. The deep C~{\sc iv} features centred at $\sim$$-$4500 and $\sim$$-$9500 km s$^{-1}$ do not show changes comparable to previous variability when going from epoch 3 to epoch 4. However the weaker feature at $\sim$$-$18\,500 km s$^{-1}$ shows a significant strengthening, with its profile becoming comparable to that of the epoch 1 observation. The absorption line centred at $\sim$$-$13\,000 km s$^{-1}$, which is very conspicuous at epoch 1 but almost unidentifiable at epochs 2 and 3, shows a slight increase in strength, however the significance of this change is only 1.08$\sigma$.

Although these changes from epoch 3 to epoch 4 are in the weakest C~{\sc iv} components and therefore do not significantly change the minimum column density estimate made in epoch 3, the localisation in velocity space of this strengthening to a range of $\sim$6000 km s$^{-1}$ is interesting. If the increase in absorption is only considered to be significant in the stronger of the two features, then the region over which the change occurs is reduced to $\sim$4600 km s$^{-1}$. A single varying trough allows the possibility of a change in the covering fraction of the absorber to explain the phenomenon. This trough is also interesting in that it does not show as great a change between epoch 2 and 3 as the feature centred at $-$9500 km s$^{-1}$. This could be due to either (a) the line being the result of a moderately saturated outflow whose peak trough depth describes the peak covering fraction, or (b) the outflow is located in a region where it is relatively shielded from the ionizing flux variations (the large velocity separation from the other troughs suggests it is in a physically distinct location). 

\section{Summary}

This following provides a summary of the conclusions that can be drawn from this investigation.\\

\noindent (i) When comparing to large-sample statistical studies of BALQSOs \citep{filizak13,filizak14,wildy14}, BALs in SDSS J1138+3517 are exceptionally variable, especially in the 360 day rest-frame time interval between epoch 1 and epoch 2. When measuring BAL variability using the method of \citet{filizak13,filizak14}, the most variable C~{\sc iv} BAL in this object undergoes a change of $\Delta$$EW$=$-$13.8 \AA{}. When compared to Fig.~29 of \citet{filizak13}, this change in $\Delta$$EW$ is far outside the range predicted by their random-walk model at an equivalent time interval. This could indicate that SDSS J1138+3517 belongs to a rare class of highly variable BALQSOs whose variability mechanism is not the same as that driving the random-walk behaviour observed in \citet{filizak13}.

\noindent (ii). When applying Gaussian fitting to model the Si~{\sc iv} outflows in Section 3.3, it is found that the components of individual doublets rarely exhibit the expected 2:1 blue to red optical depth ratio and also do not reach zero intensity at line centre. This suggests saturation of the blue component combined with spatially dependent line-of-sight optical depth (either inhomogeneous coverage or pure partial coverage).

\noindent (iii). Although both inhomogeneous power-law and PPC models successfully predict the central velocity range of the deepest saturated blue trough in epoch 2, modelling the absorption as Gaussians over the entirety of the velocity range suggests a PPC model is more successful. This is due to the fact that the best values of the power-law index suggest a very steep slope, therefore approximating a step function which is the mathematical equivalent of PPC. The fact that both models can be successfully applied over a limited range in absorption line velocity is something that should be noted in future investigations which attempt to distinguish these models.

\noindent (iv). There is no evidence for a change of covering fraction across the epochs when calculating $C$ from the (Gaussian modelled) doublets representing the two deepest absorbers, suggesting line-of-sight covering changes are not the dominant mechanism driving the absorption line variability. However the weak C~{\sc iv} feature centred at $\sim$$-$18 500 km s$^{-1}$ may have undergone a covering fraction change between epochs 3 and 4.

\noindent (v). There is evidence that ionization changes play a significant role in the variability, most notably from the coordinated changes in both Si~{\sc iv} and C~{\sc iv} absorption troughs separated widely in velocity. There is also a hint that the ionizing flux may have increased in epoch 2, given the apparent steepening of the continuum in the blue part of the spectrum. This conclusion would be supported if AGN have steeper $\alpha_{ox}$ at higher luminosities as indicated in \citet{vignali03}. The photoionization simulations are consistent with this scenario, since the column densities of both ions decrease with increasing ionization parameter in the log $U$ range spanned by the outflow. The best values for the changes in ionization parameter are within the scope of quasar EUV variability.

\noindent (vi). The location of the inner face of the outflowing gas is not well constrained, being located between approximately 0.1 and 200 parsecs from the continuum source. This range places it somewhere in a region spanning the BLR radius to narrow line region distances. However, the lowest velocity (and deepest) C~{\sc iv} trough must be absorbing both BLR and continuum emission, meaning that this outflow is located outside the BLR. Given the width of the absorption components, which are substantially less than BLR widths, and the value of $R_{BLR}$, it seems likely that the inner face is at least $\sim$1 parsec from the continuum source.  

\noindent (vii).  The magnitude of the mass outflow rate could be a substantial fraction of or even exceed the mass accretion rate. However due to the uncertainties on $R$, strong constraints on the true value of the mass outflow rate are not achieved. The derived kinetic luminosity range of $\sim$3.3$\times$10$^{42}$ to 6.6$\times$10$^{44}$ erg s$^{-1}$ indicates that feedback effects such as quenching of star formation may be possible if the true value is near the upper limit of this range.
   
\section*{Acknowledgements}

This work is supported at the University of Leicester by the Science and Technology Facilities Council (STFC) and is based on observations made with the WHT/ISIS operated on the island of La Palma by the Isaac Newton Group in the Spanish Observatorio del Roque de los Muchachos of the Instituto de Astrofísica de Canarias and spectroscopic observations from Data release 6 of the SDSS. We wish to thank our anonymous referee for their useful comments and thorough review of the draft.

Funding for the SDSS and SDSS-II has been provided by the Alfred P. Sloan Foundation, the Participating Institutions, the National Science Foundation, the U.S. Department of Energy, the National Aeronautics and Space Administration, the Japanese Monbukagakusho, the Max Planck Society, and the Higher Education Funding Council for England. The SDSS Web Site is http://www.sdss.org/.

The SDSS is managed by the Astrophysical Research Consortium for the Participating Institutions. The Participating Institutions are the American Museum of Natural History, Astrophysical Institute Potsdam, University of Basel, University of Cambridge, Case Western Reserve University, University of Chicago, Drexel University, Fermilab, the Institute for Advanced Study, the Japan Participation Group, Johns Hopkins University, the Joint Institute for Nuclear Astrophysics, the Kavli Institute for Particle Astrophysics and Cosmology, the Korean Scientist Group, the Chinese Academy of Sciences (LAMOST), Los Alamos National Laboratory, the Max-Planck-Institute for Astronomy (MPIA), the Max-Planck-Institute for Astrophysics (MPA), New Mexico State University, Ohio State University, University of Pittsburgh, University of Portsmouth, Princeton University, the United States Naval Observatory, and the University of Washington.

\bibliography{bib_cw}

\bibliographystyle{mn2e}

\bsp

\label{lastpage}

\end{document}